\documentclass[a4paper,fleqn,usenatbib]{mnras}

\usepackage[T1]{fontenc}
\usepackage{ae,aecompl}

\usepackage{graphicx}	% Including figure files
\usepackage{amsmath}	% Advanced maths commands
\usepackage{amssymb}	% Extra maths symbols
\usepackage{siunitx}

\usepackage[export]{adjustbox}

\usepackage{setspace}
\renewcommand*{\href}[3][]{#3}

\usepackage{subcaption}
\usepackage{etoolbox}
\makeatletter
\patchcmd\@combinedblfloats{\box\@outputbox}{\unvbox\@outputbox}{}{%
   \errmessage{\noexpand\@combinedblfloats could not be patched}%
}%
 \makeatother

\newcommand{\be}{\begin{equation}}
\newcommand{\ee}{\end{equation}}
\newcommand{\bea}{\begin{eqnarray}}
\newcommand{\eea}{\end{eqnarray}}

\def\Sec#1{Section~\ref{sec:#1}}

\def\Fig#1{Fig.~\ref{fig:#1}}
\def\Tab#1{Table~\ref{tab:#1}}

\def\ifm#1{\relax\ifmmode#1\else$\mathsurround=0pt #1$\fi}
\def\kms{\ifmmode\,{\rm km}\,{\rm s}^{-1}\else km$\,$s$^{-1}$\fi}

\def\Msun{\,{\rm M_{\odot}}}
\def\kcm3{\ifmmode\,{\rm K}\,{\rm cm}^{-3}\else K$\,$cm$^{-3}$\fi}

\def\Mstar{M_{\star}}

\def\MstarChab{M_{\rm \star,Chab}}

\def\MBH{M_{\rm BH}}

\newcommand{\lom}{LoM-50}
\newcommand{\him}{HiM-50}

\def\ltsima{$\; \buildrel < \over \sim \;$}
\def\simlt{\lower.5ex\hbox{\ltsima}}
\def\gtsima{$\; \buildrel > \over \sim \;$}
\def\simgt{\lower.5ex\hbox{\gtsima}}

\def\r200{r_{200}}
\def\m200{m_{200}}
\def\V200{V_{200}}
\def\M200{M_{200}}
\def\R200{R_{200}}

\newcommand{\ptm}{\vphantom}

\title[Variable IMFs with EAGLE -- II. IMF correlations]{Calibrated, cosmological hydrodynamical simulations with variable IMFs II: Correlations between the IMF and global galaxy properties}

\author[C. Barber et al.]{
Christopher Barber,$^{1}$\thanks{Email: \href{mailto:cbar@strw.leidenuniv.nl}{cbar@strw.leidenuniv.nl}}
Joop Schaye$^{1}$,
and Robert A. Crain$^{2}$
\\
$^{1}$Leiden Observatory, Leiden University, PO Box 9513, NL-2300 RA Leiden, The Netherlands\\
$^{2}$Astrophysics Research Institute, Liverpool John Moores University, 146 Brownlow Hill, Liverpool L3 5RF, UK
}

\date{Accepted XXX. Received YYY; in original form ZZZ}

\pubyear{2016}

\graphicspath{{Figures_2/}}

\begin{document}
\label{firstpage}
\pagerange{\pageref{firstpage}--\pageref{lastpage}}
\maketitle

\begin{abstract}
%This is a simple template for authors to write new MNRAS papers.
%The abstract should briefly describe the aims, methods, and main results of the paper.
%It should be a single paragraph not more than 250 words (200 words for Letters).
%No references should appear in the abstract.

The manner in which the stellar initial mass function (IMF) scales with global galaxy properties is under debate. We use two hydrodynamical, cosmological simulations to predict possible trends for two self-consistent variable IMF prescriptions that respectively become locally bottom-heavy or top-heavy in high-pressure environments. Both simulations have been calibrated to reproduce the observed correlation between central stellar velocity dispersion and the excess mass-to-light ratio (MLE) relative to a Salpeter IMF by increasing the mass fraction of, respectively, dwarf stars or stellar remnants.  We find trends of MLE with galaxy age, metallicity and [Mg/Fe] that agree qualitatively with observations. Predictions for correlations with luminosity, half-light radius, and black hole mass are presented. The significance of many of these correlations depends sensitively on galaxy selection criteria such as age, luminosity, and morphology. For an IMF with a varying high-mass end, some of these correlations are stronger than the correlation with the birth ISM pressure (the property that governs the form of the IMF), because in this case the MLE has a strong age dependence. Galaxies with large MLE tend to have overmassive central black holes. This indicates that the abnormally high MLE observed in the centres of some high-mass galaxies does not imply that overmassive BHs are merely the result of incorrect IMF assumptions, nor that excess M/L ratios are solely the result of overmassive BHs. Satellite galaxies tend to scatter toward high MLE due to tidal stripping, which may have significant implications for the inferred stellar masses of ultracompact dwarf galaxies. 

\end{abstract}


\section{Introduction}

The physical interpretation of observational diagnostics of extragalactic stellar populations, as well as predictions for such diagnostics from galaxy formation models, rely on the assumed distribution of masses of stars at birth in a given simple stellar population, the stellar initial mass function (IMF). Such studies often assume a universal functional form, motivated by the apparent universality of the IMF within the Milky Way (MW) galaxy \citep{Kroupa2001, Chabrier2003a, Bastian2010}. 

Recent evidence for IMF variations in the centres of high-mass early-type galaxies challenge this assumption of universality. Some evidence comes from dynamical studies that measure the excess central stellar mass-to-light ratio (MLE) relative to that expected given a fixed IMF. The MLE is typically measured dynamically either via gravitational lensing  \citep[e.g.][]{Auger2010, Treu2010, Spiniello2011, Barnabe2013, Sonnenfeld2015, Posacki2015, Smith2015b, Collier2018} or stellar kinematics  \citep[e.g.][]{Thomas2011, Dutton2012, Tortora2013, Cappellari2013b, Li2017}, with most studies finding larger values than one would expect for a MW-like IMF. This excess mass may come from excess dim, low-mass, dwarf stars that contribute more to the mass than the light, implying a steeper (bottom-heavy) IMF, or from stellar remnants such as black holes or neutron stars, implying a shallower (top-heavy) form. Some information about the functional form of the IMF can be inferred from spectroscopic studies, which indicate that fits to IMF-sensitive stellar absorption features require a larger ratio of dwarf to giant stars, implying that the IMF has a steeper slope either at all masses \citep[e.g.][]{Cenarro2003, vanDokkum2010, Spiniello2012, Ferreras2013, Spiniello2014}, or only at the low-mass end \citep[e.g.][]{Conroy2012b, Conroy2017} or only the high-mass end \citep[e.g.][]{Ferreras2013, LaBarbera2013, Rosani2018}. Interestingly, observations of local vigorously star-forming galaxies instead imply that the IMF becomes more top-heavy with increasing star formation rate \citep{Gunawardhana2011}.

The majority of these studies find that the IMF becomes ``heavier'' with increasing central stellar velocity dispersion, $\sigma$. To understand in more detail what drives IMF variations, it is useful to investigate how the IMF varies as a function of other galaxy properties as well. In the observational literature, there seems to be little consensus regarding the correlation between the IMF and galaxy properties other than central $\sigma$, the most notable being [Mg/Fe]. Some spectroscopic studies report a strong correlation between the MLE and [Mg/Fe] for ETGs, even stronger than that with $\sigma$ \citep{Conroy2012b, Smith2012}. On the other hand, the spectroscopic study of \citet{LaBarbera2015} concludes that, while the IMF slope correlates with both $\sigma$ and [Mg/Fe] in stacked SDSS spectra of high-$\sigma$ ETGs, the correlation with [Mg/Fe] vanishes at fixed $\sigma$. The dynamical study of \citet{McDermid2014} finds a significant (but weak) trend of MLE with [Mg/Fe] for ATLAS$^{\rm 3D}$ galaxies which however does not appear to be as strong as the correlation with $\sigma$ \citep{Cappellari2013b}. \citet{Smith2014} shows that studies that employ dynamical methods tend to favour trends between $\sigma$ with little [Mg/Fe] residual dependence, while spectroscopic methods favour a [Mg/Fe] correlation with no residual $\sigma$ dependence, even when applied to the same galaxy sample. 

The situation is even more uncertain for trends between the IMF and stellar metallicity. Spatially-resolved spectroscopic IMF studies have found that the IMF correlates strongly with local stellar metallicity \citep{Martin-Navarro2015c, Conroy2017}, while global trends tend to be weaker, with spectroscopic studies finding only weak trends \citep{Conroy2012b}, and dynamical studies finding no significant correlation at all \citep{McDermid2014, Li2017}. These discrepancies between the IMF scalings among observational IMF studies are often chalked up to differences in modelling procedures and unknown systematic biases \citep{Clauwens2015}. \citet{Clauwens2016} showed that given the uncertain observational situation, the consequences of the inferred IMF variations for the interpretation of observations of galaxy populations could vary from mild to dramatic.

Recently, \citet[][, hereafter Paper I]{Barber2018a} presented a suite of cosmological, hydrodynamical simulations that self-consistently vary the IMF on a per-particle basis as a function of the ISM pressure from which star particles are born. These simulations, which adopt respectively a bottom-heavy and a top-heavy IMF, use the EAGLE model for galaxy formation \citep{Schaye2015}. They reproduce the observed $z \approx 0$ galaxy luminosity function, half-light radii and black hole masses, and the IMF dependence on pressure has been calibrated to reproduce the observed MLE$-\sigma$ relation. The goal of this paper is to determine, for the first time, the relationships between the IMF and global galaxy properties that arise from a self-consistent, hydrodynamical, cosmological model of galaxy formation and evolution with {\it calibrated} IMF variations. In doing so, we can inform on the differences (and similarities) in such relationships as a result of differences in IMF parametrizations.

This paper is organized as follows. In \Sec{sims} we summarize the variable IMF simulations. \Sec{MLE_vs_IMF} shows the circumstances for which the MLE is a reasonable tracer of the IMF. \Sec{IMF_trends} shows the resulting correlations between the MLE and various galaxy properties, including age, metallicity, [Mg/Fe] stellar mass, luminosity, and size. \Sec{MLE_vs_MBH_Mstar} shows how galaxies with overmassive BHs tend to also have a high MLE. \Sec{most_important_param} investigates which observables most closely correlate with MLE. \Sec{satellites} examines the environmental effects on the MLE$-\sigma$ relation. We summarize in \Sec{conclusions}. In a future work (Paper III) we will discuss the spatially-resolved IMF trends within individual galaxies, including the effect of a variable IMF on radial abundance gradients, as well as on the MLE$-\sigma$ relation at high redshift. The simulation data is publicly available at \url{http://icc.dur.ac.uk/Eagle/database.php}.

\section{Simulations}
\label{sec:sims}

In this paper we investigate IMF scaling relations using cosmological, hydrodynamical simulations that self-consistently vary the IMF on a per-particle basis. These simulations were presented in Paper I, and are based on the EAGLE model \citep{Schaye2015, Crain2015, McAlpine2016}. Here we give a brief overview of EAGLE and the modifications made to self-consistently implement variable IMF prescriptions. We refer the reader to \citet{Schaye2015} and Paper I for further details.

The simulations were run using a heavily modified version of the Tree-PM smooth particle hydrodynamics (SPH) code Gadget-3 \citep{Springel2005}, on a cosmological periodic volume of (50 Mpc)$^3$ with a fiducial ``intermediate'' particle mass of $m_{\rm g} = 1.8 \times 10^6\,\Msun$ and $m_{\rm DM} = 9.7\times 10^{6}\,\Msun$ for gas and dark matter, respectively. The gravitational softening length was kept fixed at 2.66 comoving kpc prior to $z=2.8$, switching to a fixed 0.7 proper kpc thereafter. Cosmological parameters were chosen for consistency with Planck 2013 in a Lambda cold dark matter cosmogony \citep[$\Omega_{\rm b} = 0.04825$, $\Omega_{\rm m} = 0.307$, $\Omega_{\Lambda}=0.693$, $h = 0.6777$;][]{Planck2014}.

The reference EAGLE model employs analytical prescriptions to model physical processes that occur below the resolution limit of the simulation (referred to as ``subgrid'' physics). The 11 elements that are most important for radiative cooling and photoheating of gas are tracked individually through the simulation, with cooling and heating rates computed according to \citet{Wiersma2009a} subject to an evolving, homogeneous UV/X-ray background \citep{Haardt2001}. Once gas particles reach a metallicity-dependent density threshold that corresponds to the transition from the warm, atomic to the cold, molecular gas phase \citep{Schaye2004}, they become eligible for stochastic conversion into star particles at a pressure-dependent star formation rate that reproduces the Kennicutt-Schmidt star formation law \citep{Schaye2008}. Star particles represent coeval simple stellar populations that, in the reference model, adopt a \citet{Chabrier2003a} IMF. They evolve according to the lifetimes of \citet{Portinari1998}, accounting for mass loss from winds from massive stars and AGB stars, as well as supernovae (SN) types II and Ia \citep{Wiersma2009b}. Stellar ejecta are followed element-by-element and are returned to the surrounding interstellar medium (ISM), along with thermal energetic stellar feedback \citep{DallaVecchia2012} whose efficiency was calibrated to match the $z\approx 0$ galaxy stellar mass function (GSMF) and galaxy sizes. Supermassive black holes (BHs) are seeded in the central regions of high-mass dark matter haloes, and grow via accretion of low angular momentum gas \citep{Springel2005a, Booth2009, Rosas-Guevara2015} and mergers with other BHs, leading to thermal, stochastic active galactic nucleus (AGN) feedback \citep{Schaye2015} that acts to quench star formation in high-mass galaxies.

The two simulations used in this study use the same subgrid physics prescriptions as the reference EAGLE model, except that the IMF is varied as a function of the pressure of the ISM from which individual star particles form. To ensure that the simulations remain self-consistent, the stellar mass loss, nucleosynthetic element production, stellar feedback, and star formation law are all modified to be consistent with the IMF variations (see Paper I for details). The variable IMF simulations have the same volume, initial conditions, and resolution as the Ref-L050N0752 (hereafter referred to as Ref-50) simulation of \citet{Schaye2015}.

Our two variable IMF simulations differ only in their prescriptions for the IMF. In the first, which we refer to as \lom{}, the low-mass slope of the IMF (from 0.1 to $0.5\Msun$) is varied while the slope at higher masses remains fixed at the \citet{Kroupa2001} value of $-2.3$. In this prescription the IMF becomes bottom-heavy in high-pressure environments, with the slope ranging from 0 to $-3$ in low- and high-pressure environments, respectively, transitioning smoothly between the two regimes via a sigmoid function over the range $P/k_B \approx 10^4-10^6\kcm3$. Such a prescription produces stellar populations with larger stellar $M/L$ ratios at high pressures due to an excess mass fraction of low-mass dwarf stars that contribute significantly to the mass but not to the light.

For the second simulation, hereafter \him{}, we instead keep the low-mass slope fixed at the Kroupa value of $-1.3$, and vary the high-mass slope (from 0.5 to $100 \Msun$) from $-2.3$ to $-1.6$ with increasing birth ISM pressure, transitioning smoothly over the same pressure range as in the \lom{} simulation. This prescription increases the $M/L$ relative to a Kroupa IMF at high pressures by increasing the mass fraction of short-lived high-mass stars, resulting in a larger fraction of stellar remnants such as BHs, neutron stars, and white dwarfs, and lower luminosity once these high-mass stars have died off (after a few $100\, {\rm Myr}$). Note that varying the IMF with pressure is essentially equivalent to varying it with star formation rate surface density, since the latter is determined by the former in the EAGLE model.

These IMF parametrizations were individually calibrated to match the observed trend between the excess mass-to-light ratio relative to that expected for a Salpeter IMF (hereafter the MLE) and central stellar velocity dispersion found by \citet{Cappellari2013b} for high-mass elliptical galaxies. This calibration was done in post-processing of the reference (100 Mpc)$^3$ EAGLE model (Ref-L100N1504) using the Flexible Stellar Population Synthesis (FSPS) software package \citep{Conroy2009, Conroy2010}. Specifically, the allowed range of IMF slopes and the pressure range over which the IMF gradually transitions from one slope to the other were tuned until an acceptable qualitative match to the \citet{Cappellari2013b} trend was obtained. We refer the reader to Paper I for further details on the calibration procedure.  In Paper I we verified that the variable IMF runs reproduce the \citet{Cappellari2013b} trend between the MLE and velocity dispersion, but we also demonstrated that calibrating the IMF to reproduce that trend does not guarantee a match to other observational constraints on the IMF, such as the dwarf-to-giant ratio in ETGs or the ratio of ionizing to UV flux in star-forming galaxies.

In Paper I we also showed that our variable IMF simulations maintain agreement with the observables used to calibrate the EAGLE model: the present-day galaxy luminosity function, the relations between galaxy luminosity and half-light radius and black hole mass, and the global rate of type Ia SNe. This result may seem surprising given that the IMF governs the strength of stellar feedback, to which these calibration observables are quite sensitive (Crain et al. 2015). The fact that these galaxy observables are not strongly affected by the modified stellar feedback is likely due to the following (simplified) picture, which we separate into star-forming and quenched regimes: 

If galaxy formation is self-regulated and if the outflow rate is large compared with the star formation rate, then the outflow rate will tend to adjust to balance the inflow rate when averaged over sufficiently long timescales. If we neglect preventative feedback and recycling, then the gas inflow rate tracks that of the dark matter and does not depend on the IMF. If the IMF is modified, then a star-forming galaxy of fixed mass will adjust its star formation rate (SFR) to ensure that the same feedback energy is released in order to generate the same outflow rate that is needed to balance the inflow rate. For a top-heavy IMF, galaxies need to form fewer stars relative to the case of a standard IMF to obtain the same feedback energy. This results in lower SFRs, and thus lower ratios of stellar mass to halo mass, resulting in a lower normalization of the GSMF. However, for star-forming galaxies, $M/L$ is lower due to the top-heavy IMF, so the luminosity at fixed halo mass ends up being similar to the Chabrier case. According to \citet{Booth2010} and \citet{Bower2017}, BH mass is a function of halo mass (for sufficiently large halo masses) for a fixed AGN feedback efficiency, so the $\MBH-L$ relation is also not strongly affected. For a bottom-heavy IMF, this situation is reversed, where more stars are required to obtain the same feedback energy, increasing the GSMF, but their higher $M/L$ ratios (due to an increased fraction of dwarf stars) makes the luminosity function (and the $\MBH-L$ relation) similar to the Chabrier case. 

For low-mass galaxies these effects are small in our simulations, since the IMF only varies away from Chabrier at the high pressures typical of high-mass galaxies. In this regime, AGN feedback quenches galaxies at a particular virial temperature \citep[or rather entropy;][]{Bower2017}, leading to an approximately fixed BH mass -- halo mass relation (because $\MBH$ must be sufficiently high to drive an outflow and quench star formation). For a top-heavy IMF, the lower $\Mstar/M_{200}$ (assuming the stellar mass formed when star-forming) leads to higher $\MBH/\Mstar$ and a lower GSMF. A quenched galaxy with a top-heavy IMF can have a higher or lower $M/L$ depending on how long it has been quenched -- if quenched for more than $\approx 3$ Gyr, $M/L$ is higher so the luminosity function cuts off at lower luminosity. However, since high-mass galaxies are not as strongly quenched in \him{}, this effect is small. For a bottom-heavy IMF, everything is reversed: galaxies are quenched at higher $\Mstar$, leading to a higher GSMF, but since $M/L$ is higher, they quench at lower luminosity so the luminosity function remains similar. 

In the intermediate mass regime (around the knee of the GSMF), the situation is more complex since both stellar feedback and AGN feedback play an important role in self-regulation and the star formation law becomes important (see Paper I). As noted above, a top-heavy IMF leads to a lower SFR because of the larger amount of feedback energy per unit stellar mass formed. This would imply a lower gas fraction, which would reduce the BH growth. However, this effect is counteracted by the decreased normalization of the observed star formation law at high pressures (relative to that of a standard IMF), which increases the gas surface densities at fixed star formation rate surface density. Thus, AGN feedback and BH growth are not strongly affected at fixed halo mass and galaxy luminosity. Again, the situation is reversed for bottom-heavy IMF variations. These effects, as well as the relatively poor statistics at the high-mass end relative to the (100 Mpc)$^3$ EAGLE simulation, likely eliminated any need to adjust the feedback parameters originally used to calibrate the EAGLE model.

Structures in the simulation are separated into ``haloes'' using a friends-of-friends halo finder with a linking length of 0.2 times the mean inter-particle spacing \citep{Davis1985}. Galaxies are identified within haloes as self-bound structures using the SUBFIND algorithm \citep{Springel2001,Dolag2009}. We consider only galaxies with at least 500 stellar particles, corresponding to a stellar mass $\Mstar \approx 9 \times 10^{8} \Msun$. Galaxies in the mass range of interest in this study are sufficiently well resolved, as those with $\sigma_e > 80\kms$ have $\Mstar > 10^{10} \Msun$, corresponding to $\gtrsim 5600$ stellar particles. Unless otherwise specified, all global galactic properties shown in this paper (e.g. MLE, age, metallicity, [Mg/Fe]) are computed considering star particles within the 2d projected half-light radius, $r_e$, of each galaxy, measured with the line-of-sight parallel to the $z$-axis of the simulation box. 

\section{Is the $(M/L)$-excess a good tracer of the IMF?}
\label{sec:MLE_vs_IMF}

%fig 1: IMF slope vs MLE
\begin{figure*}
\includegraphics[width=\textwidth]{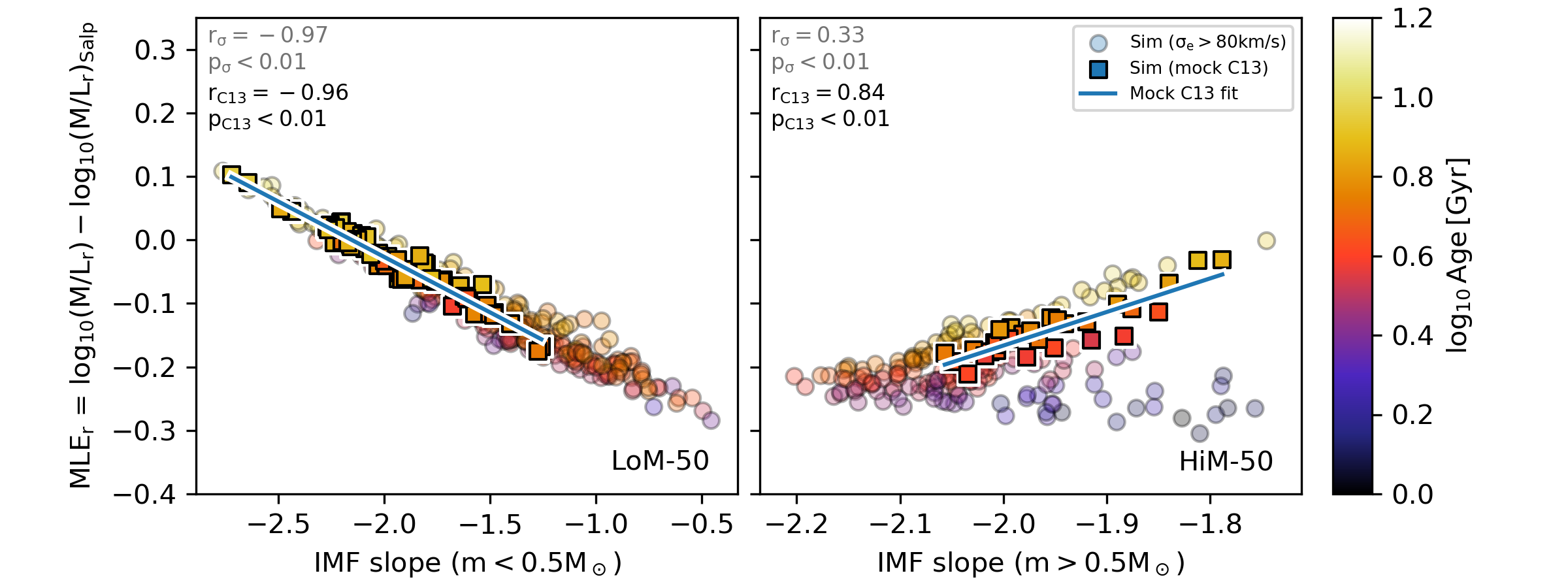}

\caption{ Excess $r$-band mass-to-light ratio relative to that for a Salpeter IMF as a function of IMF slope for galaxies with $\sigma_e > 10^{1.9}\ (\approx 80)\kms$ at $z=0.1$, coloured by stellar age. All quantities are $r$-band light-weighted means measured within the 2d projected $r$-band half-light radius. The left and right panels show low-mass ($m < 0.5 \Msun$) and high-mass ($m > 0.5 \Msun$) IMF slopes for \lom{} and \him{}, respectively. Galaxies selected in a similar way to C13 (see text) are shown as opaque squares while all others are shown as translucent circles. The Pearson correlation coefficient, $r$, and its $p$-value are indicated in each panel for the $\sigma_e > 10^{1.9}\kms$ and mock C13 samples in grey and black, respectively. Blue solid lines show least-squares fits to the mock C13 samples (see \Tab{fits}). MLE$_r$ is an excellent proxy for the low-mass IMF slope variations, but is only a good proxy for high-mass slope variations for old galaxies, with age-dependent scatter.}
\label{fig:MLE_vs_IMFslopes}
\end{figure*}

We wish to investigate trends between the IMF and global galaxy properties in a way that is testable with observations. Since dynamical studies use the MLE as a proxy for the IMF, it is important to check that this parameter correlates with the IMF for our galaxy sample. For each galaxy, we compute the MLE relative to the Salpeter IMF as 
\be
{\rm MLE}_r = \log_{10}(M/L_r) - \log_{10}(M/L_r)_{\rm Salp},
\label{eqn:MLE}
\ee
where $M$ and $L_r$ are the true stellar mass and SDSS $r$-band luminosity of each galaxy, respectively, and  $(M/L_r)_{\rm Salp}$ is the stellar mass-to-light ratio that the galaxy would have had if evolved with a Salpeter IMF given the same distribution of ages, initial masses, and metallicities of its stars (note that $(M/L_r)_{\rm Salp}$ is equivalent to the ratio between the Salpeter-inferred stellar mass [$L_r \times (M/L_r)_{\rm Salp}$] and the true luminosity). Luminosities and masses of individual star particles are computed using FSPS\footnote{We use the Basel spectral library \citep{Lejeune1997, Lejeune1998, Westera2002} with Padova isochrones \citep{Marigo2007, Marigo2008}.}, given each star particle's age, metallicity, and IMF. We make no dust correction other than ignoring the luminosities of star particles with ${\rm age} < 10\, {\rm Myr}$, as such stars are expected to be obscured by their birth clouds \citep[e.g.][]{Charlot2000}. 

In \Fig{MLE_vs_IMFslopes} we show MLE$_r$ as a function of IMF slope for galaxies with $\sigma_e > 80\kms$ in our variable IMF simulations at $z=0.1$, coloured by age. For \lom{}, MLE$_r$ is an excellent tracer of the IMF, with very little dependence on age or metallicity.  For \him{}, MLE$_r$ is only a good tracer of the IMF at fixed age (and, ideally, old age since the slope is very shallow for ages $\lesssim 3\,$Gyr), as can be seen by the strong vertical age gradient (i.e. the colour of the data points) at fixed high-mass slope in the right panel. 

To compare our results with observed trends between MLE$_r$ and galaxy properties in the literature, we select galaxies from our simulations using approximately the same selection criteria as the ATLAS$^{\rm 3D}$ sample used by \citet[][hereafter C13]{Cappellari2013b}. Their galaxy sample is complete down to $M_{\rm K} = -21.5$ mag, consisting of 260 morphologically-selected elliptical and lenticular galaxies, chosen to have old stellar populations (H$\beta$ equivalent width less than $2.3 \buildrel _\circ \over {\mathrm{A}}$). For our ``mock C13'' sample, we select galaxies with $M_{\rm K} < -21.5$ mag and intrinsic $u^*-r^* > 2$. The $u^*-r^*$ colour cut roughly separates galaxies in the red sequence from the blue cloud for EAGLE galaxies \citep{Correa2017} and ensures that we exclude galaxies with light-weighted ages younger than $\approx 3$ Gyr.

The mock C13 galaxies are highlighted as opaque squares in \Fig{MLE_vs_IMFslopes}. Since these galaxies are selected to be older than $\approx 3\, {\rm Gyr}$, their MLE is a reasonable tracer of the IMF in \him{}, but with more scatter than for \lom{} due to the residual age dependence.\footnote{Note that the trend flattens at the Kroupa value for high-mass slopes steeper than $-2.1$, and the trend would actually reverse if the high-mass slope were to become steeper than Salpeter ($< -2.35$).} These dependencies should be kept in mind when interpreting the trends between the MLE$_r$ and the global galaxy properties shown in the next section.

\section{Trends between the $(M/L)$-excess and global galactic properties}
\label{sec:IMF_trends}
\begin{figure*}

\includegraphics[width=\textwidth]{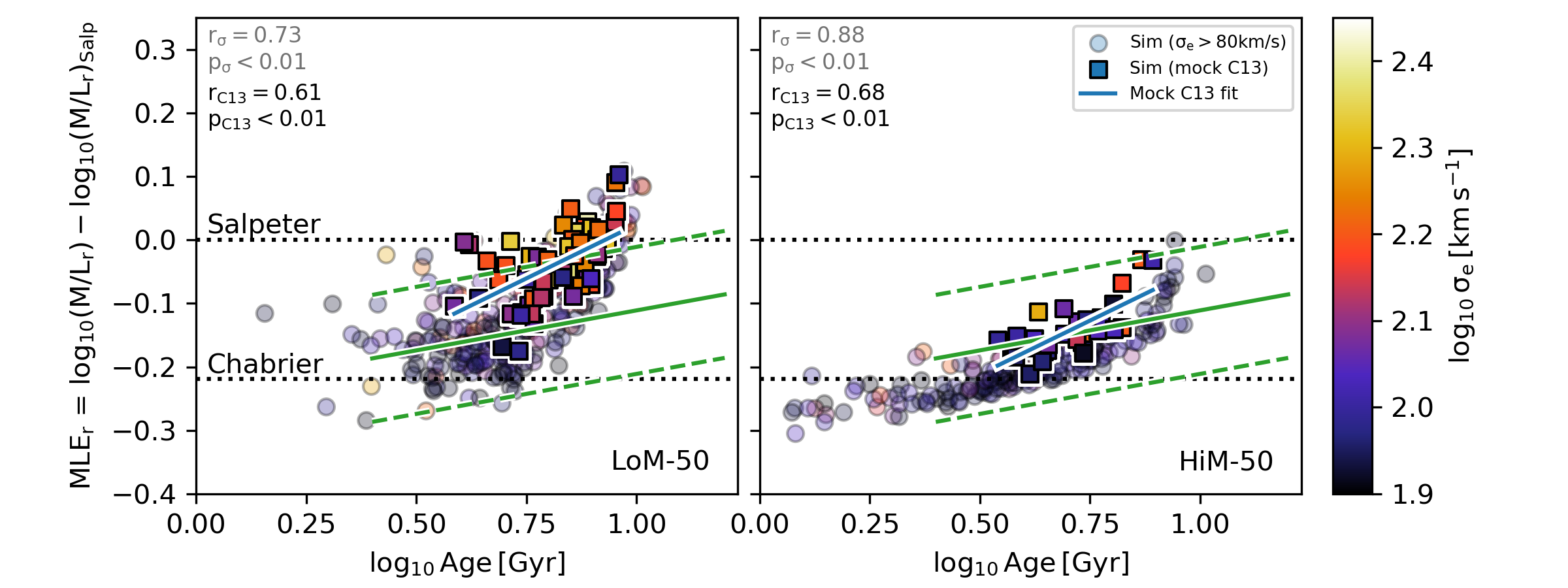}
\includegraphics[width=\textwidth]{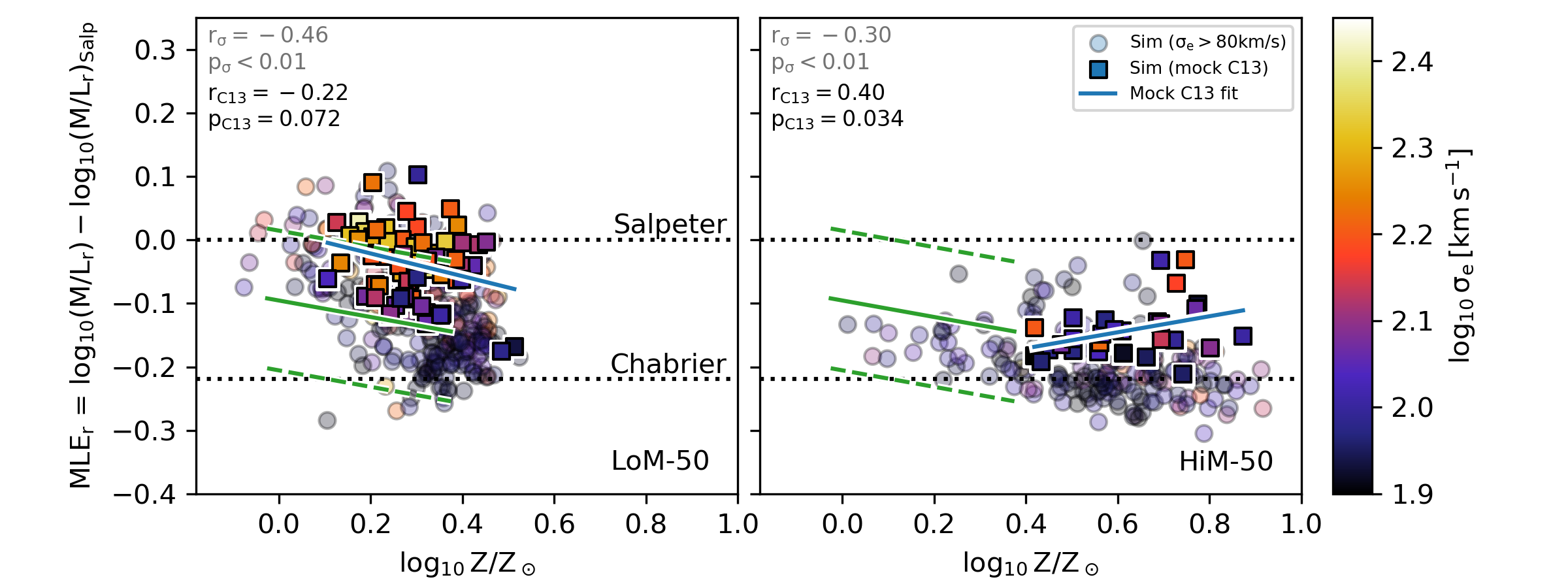}
\includegraphics[width=\textwidth]{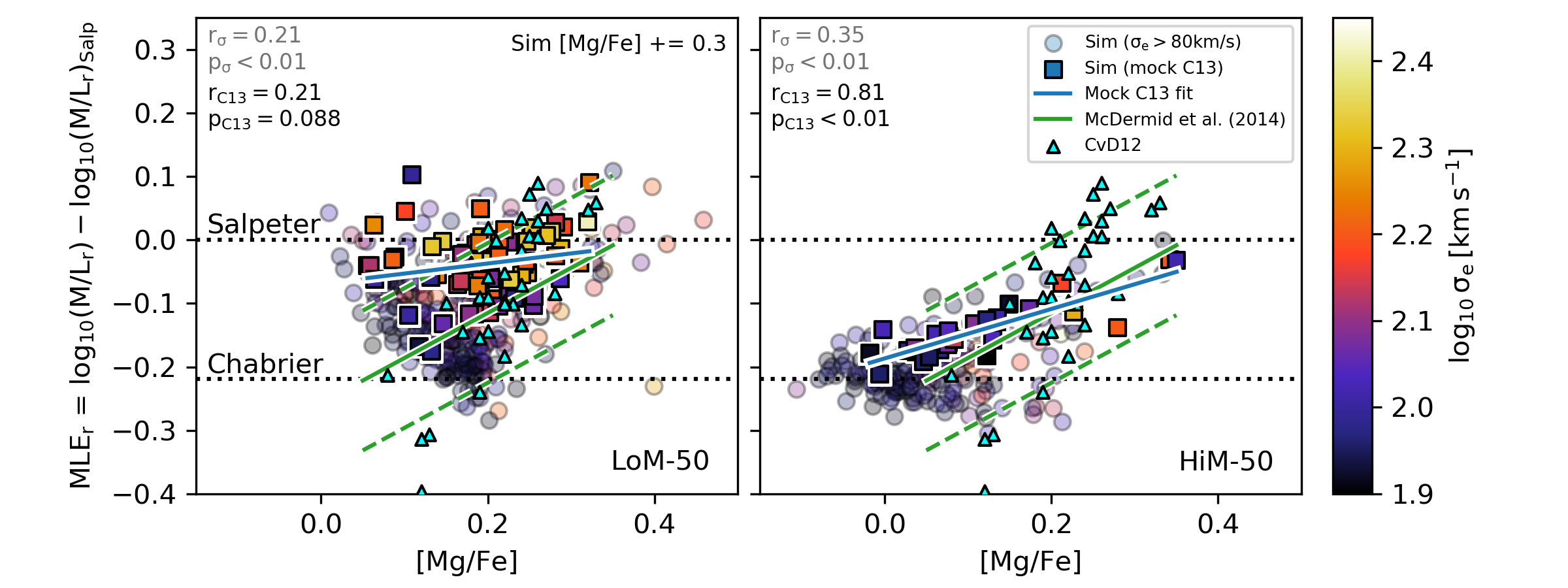}

\caption{Excess $r$-band $M/L$-ratio with respect to that for a Salpeter IMF (MLE$_r$) as a function of $r$-band light-weighted mean age (top row), stellar metallicity (middle row) and stellar alpha-enhancement (bottom row), for galaxies with $\sigma_e > 10^{1.9}\ (\approx 80)\kms$ in the \lom{} (left column) and \him{} (right column) simulations at $z=0.1$. Points are coloured by $\sigma_e$. All quantities are computed within (2d-projected) $r_e$. Galaxies selected in a similar way to C13 (see text) are shown as opaque squares while all others are shown as translucent circles. Blue solid lines show least-squares fits to the mock C13 samples (see \Tab{fits}). The simulation values for [Mg/Fe] have been increased by 0.3 dex for \lom{}. We assume $Z_\odot = 0.127$ and take other solar abundances from \citet{Asplund2009}.   We show the linear fits with $1 \sigma$ scatter found in \citet{McDermid2014} as green-solid and -dashed lines, respectively, and the MLE$-$[Mg/Fe] relation of \citet{Conroy2012b} as cyan triangles. Metallicities from \citet{McDermid2014} have been converted to our solar scale. For both variable IMF simulations we see strong positive correlations of MLE$_r$ with age and [Mg/Fe]. When considering all galaxies with $\sigma_e > 80\kms$, we find a weak but significant negative correlation with $Z$ for both simulations, but the correlation disappears for mock C13 galaxies.}
\label{fig:IMF_vs_global_props}
\end{figure*}

There is currently much debate regarding possible trends between the IMF and global galaxy properties other than velocity dispersion, such as age, metallicity, and alpha enhancement. In this section we investigate these trends in our (self-consistent, calibrated) variable IMF simulations, where the IMF is governed by the local pressure in the ISM. In the left and right columns of \Fig{IMF_vs_global_props} we show MLE$_r$ as a function of these properties in \lom{} and \him{}, respectively, and compare with observed trends for ETGs from the ATLAS$^{\rm 3D}$ survey \citep{McDermid2014} and another sample from \citet{Conroy2012b}. Note that while \citet{McDermid2014} measure MLE within a circular aperture of radius $r_e$, they \citep[as well as ][]{Conroy2012b}\footnote{Also note that \citet{Conroy2012b} measure dynamical $M/L$ within $r_e$ but spectroscopic $M/L$ within $r_e/8$.} measure age, metallicity, and [Mg/Fe] within $r_e/8$. For many of our galaxies, $r_e/8$ would be close to the gravitational softening scale of the simulation. Indeed, performing our analysis within this aperture only serves to add resolution-related noise to the plots. Thus, since \citet{McDermid2014} claim that their results are unchanged for an aperture choice of $r_e$, we report our results consistently within $r_e$. For completeness we show all galaxies with $\sigma_e > 10^{1.9}\ (\approx 80)\kms$ (the $\sigma_e$-complete sample; translucent circles), but focus our comparison with observations on galaxies consistent with the C13 selection criteria (opaque squares). This $\sigma_e$ limit of $80\kms$ was chosen because it is the lowest $\sigma_e$ value that our mock C13 samples reach.

%fig 2: IMF vs age, met, and MgFe

\begin{table*}
\caption{Fit parameters to the $r$-band mass-to-light ratio excess, MLE$_r$, for mock C13 galaxies in our variable IMF simulations. Columns 2 and 3 show the result of a linear least-squares fit of the relation between each parameter (indicated in Column 1) individually and MLE$_r$, of the form MLE$_r = ax+b$. Column 4 shows the coefficient of determination, while columns 5 and 6 show the Spearman-$r$ value and the corresponding $p$-value.}
\begin{tabular}{lrrrrr}
$x$ & $a$ & $b$ & $R^2$ & Spearman $r$ & $p$ \\
\hline
\lom{}\\
\hline
$\log_{10} P_{\rm birth}/(\kcm3)$ & $0.12\pm0.00$ & $-0.69\pm0.02$ & $0.96$ & $0.99$ & $< 0.01$ \\
$\log_{10}\sigma_e/(\kms)$ & $0.22\pm0.05$ & $-0.52\pm0.11$ & $0.22$ & $0.48$ & $< 0.01$ \\
$\log_{10} {\rm Age/Gyr}$ & $0.34\pm0.06$ & $-0.32\pm0.05$ & $0.32$ & $0.61$ & $< 0.01$ \\
$\log_{10} Z/Z_\odot$ & $-0.18\pm0.08$ & $0.01\pm0.02$ & $0.08$ & $-0.22$ & $0.07$ \\
$[$Mg/Fe$]$ & $0.16\pm0.10$ & $-0.02\pm0.01$ & $0.04$ & $0.21$ & $0.09$ \\
$\log_{10} r_e/{\rm kpc}$ & $-0.04\pm0.04$ & $-0.01\pm0.02$ & $0.02$ & $-0.06$ & $0.65$ \\
$\log_{10} M_{\rm \star,Chab}$ & $0.08\pm0.03$ & $-0.92\pm0.29$ & $0.12$ & $0.38$ & $< 0.01$ \\
$M_K - 5\log_{10}h$ & $-0.02\pm0.01$ & $-0.53\pm0.25$ & $0.06$ & $-0.26$ & $0.03$ \\
$\log_{10} M_{\rm BH}/M_\star$ & $0.11\pm0.02$ & $0.31\pm0.06$ & $0.34$ & $0.46$ & $< 0.01$ \\
$\log_{10} \sigma_e^2/r_e\ [{\rm km^2\,s^{-2}\,kpc^{-1}}]$ & $0.15\pm0.02$ & $-0.59\pm0.09$ & $0.37$ & $0.56$ & $< 0.01$ \\
\hline
\him{}\\
\hline
$\log_{10} P_{\rm birth}/(\kcm3)$ & $0.07\pm0.01$ & $-0.51\pm0.03$ & $0.87$ & $0.91$ & $< 0.01$ \\
$\log_{10}\sigma_e/(\kms)$ & $0.19\pm0.08$ & $-0.52\pm0.15$ & $0.18$ & $0.48$ & $< 0.01$ \\
$\log_{10} {\rm Age/Gyr}$ & $0.34\pm0.06$ & $-0.38\pm0.04$ & $0.53$ & $0.68$ & $< 0.01$ \\
$\log_{10} Z/Z_\odot$ & $0.13\pm0.06$ & $-0.22\pm0.04$ & $0.15$ & $0.39$ & $0.04$ \\
$[$Mg/Fe$]$ & $0.39\pm0.05$ & $-0.19\pm0.01$ & $0.72$ & $0.81$ & $< 0.01$ \\
$\log_{10} r_e/{\rm kpc}$ & $0.12\pm0.08$ & $-0.24\pm0.06$ & $0.09$ & $0.24$ & $0.21$ \\
$\log_{10} M_{\rm \star,Chab}$ & $0.09\pm0.04$ & $-1.10\pm0.40$ & $0.17$ & $0.36$ & $0.06$ \\
$M_K - 5\log_{10}h$ & $-0.02\pm0.02$ & $-0.66\pm0.34$ & $0.08$ & $-0.21$ & $0.26$ \\
$\log_{10} M_{\rm BH}/M_\star$ & $-0.01\pm0.02$ & $-0.16\pm0.06$ & $< 0.01$ & $0.24$ & $0.21$ \\
$\log_{10} \sigma_e^2/r_e\ [{\rm km^2\,s^{-2}\,kpc^{-1}}]$ & $0.08\pm0.05$ & $-0.40\pm0.15$ & $0.10$ & $0.36$ & $0.06$ \\
\end{tabular}
\label{tab:fits}
\end{table*}

\begin{table*}
\caption{As in \Tab{fits} but for all galaxies with $\sigma_e > 10^{1.9}\kms$. }
\begin{tabular}{lrrrrr}
$x$ & $a$ & $b$ & $R^2$ & Spearman $r$ & $p$ \\
\hline
\lom{}\\
\hline
$\log_{10} P_{\rm birth}/(\kcm3)$ & $0.12\pm0.00$ & $-0.70\pm0.01$ & $0.96$ & $0.98$ & $< 0.01$ \\
$\log_{10}\sigma_e/(\kms)$ & $0.28\pm0.03$ & $-0.68\pm0.07$ & $0.18$ & $0.45$ & $< 0.01$ \\
$\log_{10} {\rm Age/Gyr}$ & $0.37\pm0.02$ & $-0.37\pm0.02$ & $0.46$ & $0.72$ & $< 0.01$ \\
$\log_{10} Z/Z_\odot$ & $-0.32\pm0.03$ & $0.00\pm0.01$ & $0.22$ & $-0.47$ & $< 0.01$ \\
$[$Mg/Fe$]$ & $0.28\pm0.06$ & $-0.06\pm0.01$ & $0.07$ & $0.21$ & $< 0.01$ \\
$\log_{10} r_e/{\rm kpc}$ & $-0.14\pm0.02$ & $-0.01\pm0.01$ & $0.17$ & $-0.39$ & $< 0.01$ \\
$\log_{10} M_{\rm \star,Chab}$ & $0.04\pm0.01$ & $-0.57\pm0.15$ & $0.03$ & $0.14$ & $0.01$ \\
$M_K - 5\log_{10}h$ & $0.00\pm0.01$ & $-0.03\pm0.12$ & $< 0.01$ & $0.04$ & $0.45$ \\
$\log_{10} M_{\rm BH}/M_\star$ & $0.13\pm0.01$ & $0.32\pm0.03$ & $0.41$ & $0.68$ & $< 0.01$ \\
$\log_{10} \sigma_e^2/r_e\ [{\rm km^2\,s^{-2}\,kpc^{-1}}]$ & $0.20\pm0.01$ & $-0.82\pm0.04$ & $0.50$ & $0.72$ & $< 0.01$ \\
\hline
\him{}\\
\hline
$\log_{10} P_{\rm birth}/(\kcm3)$ & $0.05\pm0.01$ & $-0.38\pm0.03$ & $0.23$ & $0.39$ & $< 0.01$ \\
$\log_{10}\sigma_e/(\kms)$ & $0.06\pm0.04$ & $-0.28\pm0.09$ & $< 0.01$ & $0.12$ & $0.10$ \\
$\log_{10} {\rm Age/Gyr}$ & $0.22\pm0.01$ & $-0.30\pm0.01$ & $0.68$ & $0.87$ & $< 0.01$ \\
$\log_{10} Z/Z_\odot$ & $-0.08\pm0.02$ & $-0.12\pm0.01$ & $0.07$ & $-0.30$ & $< 0.01$ \\
$[$Mg/Fe$]$ & $0.32\pm0.04$ & $-0.19\pm0.00$ & $0.27$ & $0.36$ & $< 0.01$ \\
$\log_{10} r_e/{\rm kpc}$ & $0.07\pm0.02$ & $-0.22\pm0.02$ & $0.03$ & $0.28$ & $< 0.01$ \\
$\log_{10} M_{\rm \star,Chab}$ & $-0.05\pm0.01$ & $0.34\pm0.13$ & $0.07$ & $-0.24$ & $< 0.01$ \\
$M_K - 5\log_{10}h$ & $0.03\pm0.00$ & $0.42\pm0.09$ & $0.18$ & $0.39$ & $< 0.01$ \\
$\log_{10} M_{\rm BH}/M_\star$ & $0.05\pm0.01$ & $-0.03\pm0.02$ & $0.21$ & $0.52$ & $< 0.01$ \\
$\log_{10} \sigma_e^2/r_e\ [{\rm km^2\,s^{-2}\,kpc^{-1}}]$ & $-0.02\pm0.02$ & $-0.11\pm0.06$ & $< 0.01$ & $-0.09$ & $0.19$ \\
\end{tabular}
\label{tab:fits_sig1p9}
\end{table*}

\subsection{MLE vs age}

First we investigate the relationship between the MLE and galaxy age. In the top row of \Fig{IMF_vs_global_props} we show MLE$_r$ as a function of the $L_r$-weighted mean stellar age of galaxies in \lom{} (left panel) and \him{} (right panel). For both simulations we see a strong trend of increasing MLE$_r$ with age, where older galaxies tend to have higher (i.e. heavier) MLE$_r$. This result is in qualitative agreement with \citet{McDermid2014}, but with a steeper slope for \lom{}, and smaller scatter for \him{}. Note as well that the positive trend found by \citet{McDermid2014} is sensitive to the methodology used, and in fact disappears when $(M/L_r)_{\rm Salp}$ is derived from individual line strengths rather than full spectral fitting. It is thus quite interesting that we find strong positive correlations with age for both simulations. 

For \lom{}, this trend is driven by the higher pressures at which stars form at higher redshift (as was shown for Ref-50 by \citealt{Crain2015} and will be investigated further for our simulations in Paper III), yielding more bottom-heavy IMFs for older ages. For \him{}, the trend with age is tighter due to the fact that, in addition to the trend of higher birth ISM pressure with increasing formation redshift, the MLE$_r$ increases with age for a stellar population with a top-heavy IMF, even if the IMF is fixed. The result is that, even though the IMF itself depends only on birth ISM pressure, in the end the MLE$_r$ correlates more strongly with age than with IMF slope (compare the upper-right panel of \Fig{IMF_vs_global_props} with the right panel of \Fig{MLE_vs_IMFslopes}; also Fig. 2 of Paper I). This is especially true for the $\sigma_e$-complete sample due to its large range of ages, while the age dependence is reduced for the mock C13 sample due to the exclusion of young galaxies. 

We also note that galaxies with $\sigma_e>80\kms$ in \him{} extend to quite young ages ($<1$ Gyr), whereas in the \lom{} case only a handful of galaxies are $< 3$ Gyr old (although the $u^*-r^*$ selection criterion removes all galaxies with light-weighted ages younger than 3 Gyr in our mock C13 samples). This age difference is due to the higher SFRs in the \him{} simulation relative to \lom{}, which were discussed in Paper I.  We remind the reader that we ignore the luminosities of star particles with ages less than 10 Myr. Without this cut, the age and MLE$_r$ would extend to even lower values due to the ongoing star formation in \him{} galaxies.

We conclude that both \lom{} and \him{} agree qualitatively with the positive trend of MLE$_r$ with age inferred from the ATLAS$^{3D}$ survey by \citet{McDermid2014}, but with a stronger correlation. We encourage other observational IMF studies to measure the correlation between IMF diagnostics and age as well in order to help test these predictions.

\subsection{MLE vs metallicity}

Observationally, evidence for trends between the IMF and metal abundances have been reported, but with conflicting results. While spectroscopic studies find strong positive trends of bottom-heaviness with local metallicity \citep{Martin-Navarro2015c, Conroy2017}, dynamical studies find no significant correlation between MLE$_r$ and global metallicity \citep{McDermid2014, Li2017}. In the middle row of \Fig{IMF_vs_global_props} we plot MLE$_r$ vs (dust-free) $L_r$-weighted metallicity measured within $r_e$. We assume $Z_\odot = 0.0127$ and convert observationally-derived metallicities from the literature to this scale. The offset of \him{} galaxies toward higher metallicities is due to the higher nucleosynthetic yields resulting from a top-heavy IMF, as discussed in Paper I. 

For both variable IMF simulations, we see a weak but significant trend of decreasing MLE$_r$ with metallicity for the sample with $\sigma > 80\kms$. For both \lom{} and \him{}, the negative correlation of MLE$_r$ with metallicity is a consequence of the positive and negative relationships of the MLE$_r$ and $Z$, respectively, with age. Interestingly, in \him{} the negative correlation with metallicity is weaker than in \lom{} despite the stronger dependence of MLE$_r$ on age. This is the result of the strong effect that a top-heavy IMF has on the metal yields. At fixed age, a galaxy with higher MLE has on average an IMF with a shallower high-mass slope, resulting in higher metal yields and thus higher metallicities. The effect is strongest for the oldest galaxies where the scatter in MLE$_r$ is greatest (see upper right panel of \Fig{IMF_vs_global_props}). Thus, the oldest galaxies with high MLE$_r$ that would have had low metallicity with a Chabrier IMF get shifted toward higher metallicity in the middle-right panel of \Fig{IMF_vs_global_props}, reducing the strength of the negative MLE$_r-Z$ correlation. 

Restricting our sample to the mock C13 galaxies, the negative trend with $Z$ is weaker and no longer significant for \lom{}, and is weakly positive for \him{}.  The negative trend for \him{} disappears for this sample because we remove the low-metallicity, old galaxies with intermediate MLE that help drive the negative trend in the $\sigma_e$-complete selection. These galaxies are excluded from the mock C13 selection due to the luminosity cut. The weakness of these trends is in agreement with the weakly negative but non-significant correlation of \citet{McDermid2014}, as well as the lack of correlation found by \citet{Li2017}. Interestingly, our results are in stark contrast with the positive correlation between the IMF slope and spatially-resolved {\it local} local metallicity of \citet{Martin-Navarro2015c}. We make a fairer comparison to their result in Paper III, where we show that locally we do in fact find a positive correlation between MLE$_r$ and metallicity. 

We conclude that, for a $\sigma_e$-complete sample of galaxies, the MLE$_r$ is predicted to anticorrelate with total stellar metallicity for low-mass IMF slope variations, while being relatively insensitive to metallicity for high-mass slope variations. In both cases these correlations disappear for samples consistent with the selection criteria of the ATLAS$^{\rm 3D}$ survey, in agreement with dynamical studies.

\subsection{MLE vs [Mg/Fe]}
\label{sec:IMF_vs_MgFe}

In the bottom row of \Fig{IMF_vs_global_props} we plot MLE$_r$ as a function of [Mg/Fe]\footnote{Solar abundances for Mg and Fe are taken from \citet{Asplund2009}.}, both measured within $r_e$.  For \lom{} we increase the [Mg/Fe] values by 0.3 dex to facilitate comparison with the observed trends. This procedure is somewhat arbitrary, but is motivated by the fact that \citet{Segers2016} showed that [Mg/Fe] is underestimated by EAGLE and that the nucleosynthetic yields are uncertain by about a factor of 2 \citep{Wiersma2009b}; thus the slopes of these trends are more robust than the absolute values. This procedure is not necessary for \him{} due to the increased metal production resulting from a top-heavy IMF (see Fig. 9 of Paper I).

For both simulations, for the $\sigma_e>80\kms$ selection we see weak but significant positive correlations between MLE$_r$ and [Mg/Fe]. When selecting only mock C13 galaxies, the trend for \lom{} is still positive but no longer significant. These trends are in qualitative agreement with the positive trends found by \citet{Conroy2012b} and \citet{McDermid2014}, and highlight the importance of sample selection in determining the significance of these correlations. For \him{}, the MLE$_r-$[Mg/Fe] relation is stronger than in \lom{}, likely due to a combination of the fact that [Mg/Fe] correlates strongly with age in high-mass galaxies \citep{Segers2016} and the fact that the [Mg/Fe] ratios are strongly affected by the shallow high-mass IMF slopes resulting from the HiM IMF parametrization (see Paper I). The correlation strengthens for the mock C13 selection due to the exclusion of young, blue galaxies with intermediate [Mg/Fe] values, although given the low number of high-[Mg/Fe] mock C13 galaxies in \him{}, the strength of this correlation may be sensitive to our mock C13 selection criteria.

In \Sec{most_important_param} we show that the trend between MLE$_r$ and [Mg/Fe] for \lom{} galaxies is solely due to the correlation between [Mg/Fe] and $\sigma_e$, while that for \him{} is partially due to the correlation between [Mg/Fe] and age. The former is consistent with \citet{LaBarbera2015} who found that, while the observations are consistent with a correlation between MLE and [Mg/Fe], this correlation disappears at fixed central velocity dispersion.

\subsection{MLE vs Chabrier-inferred galaxy mass, luminosity, and size}

\begin{figure*}

\includegraphics[width=\textwidth]{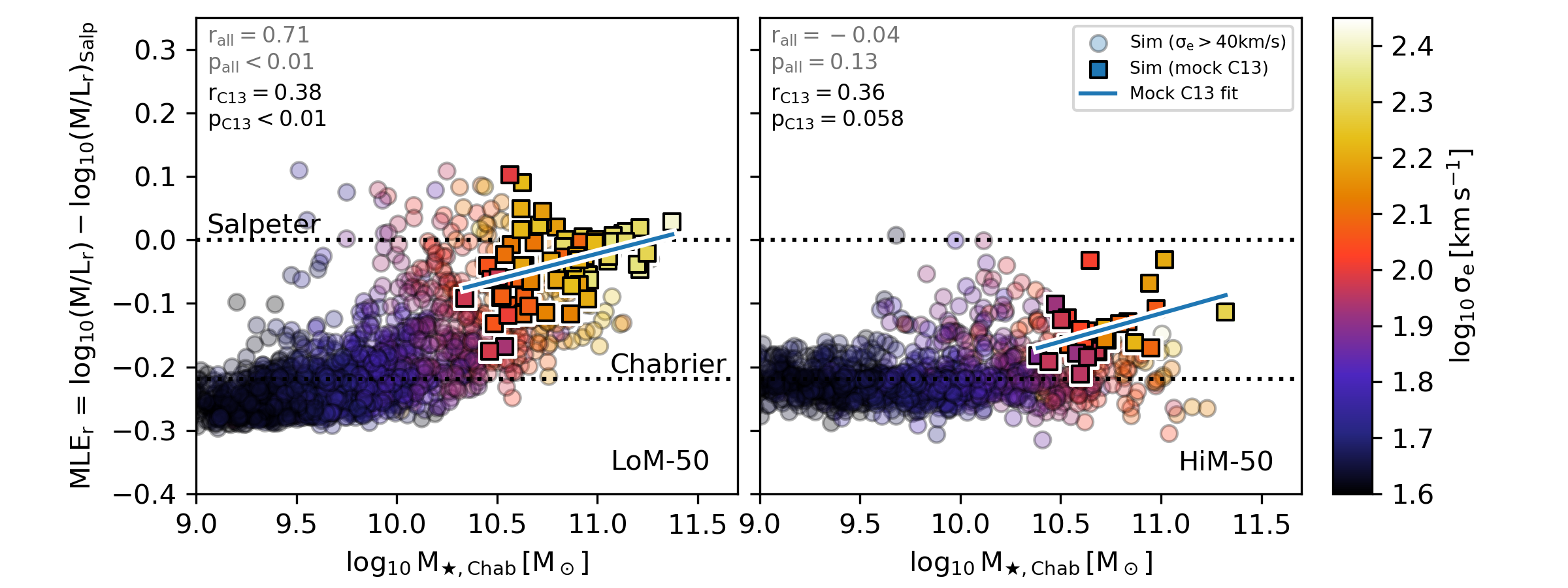}
\includegraphics[width=\textwidth]{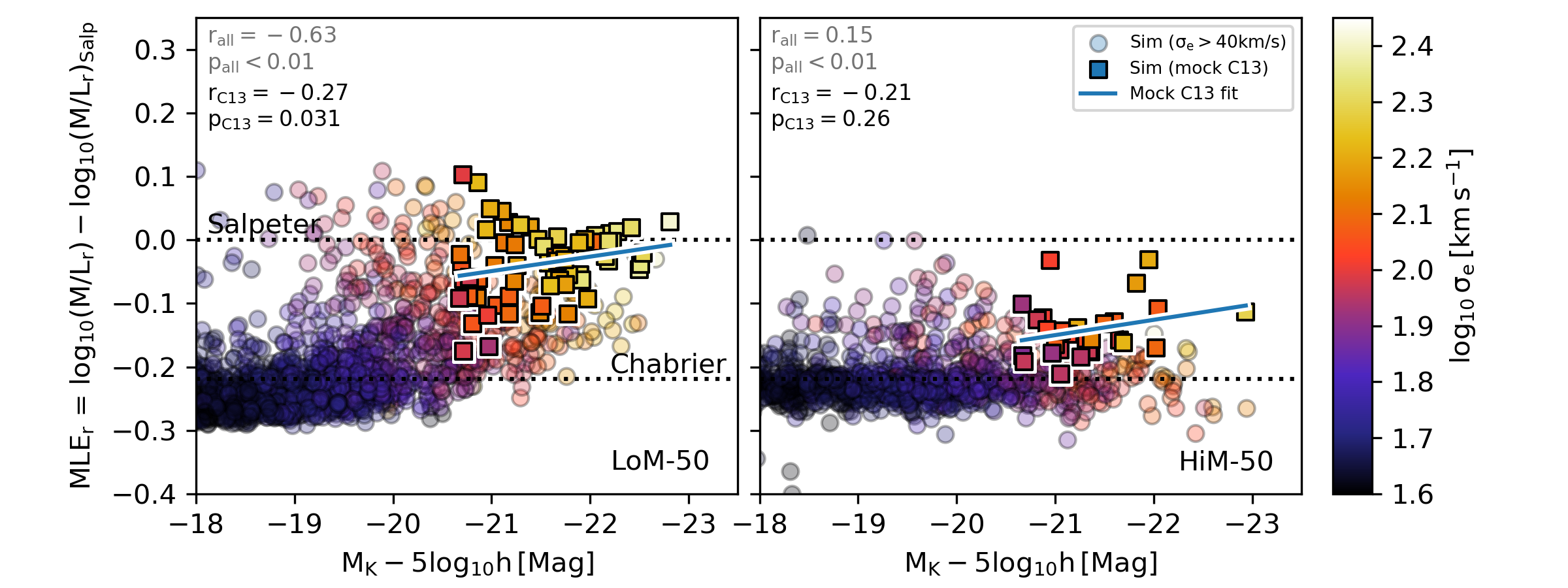}
\includegraphics[width=\textwidth]{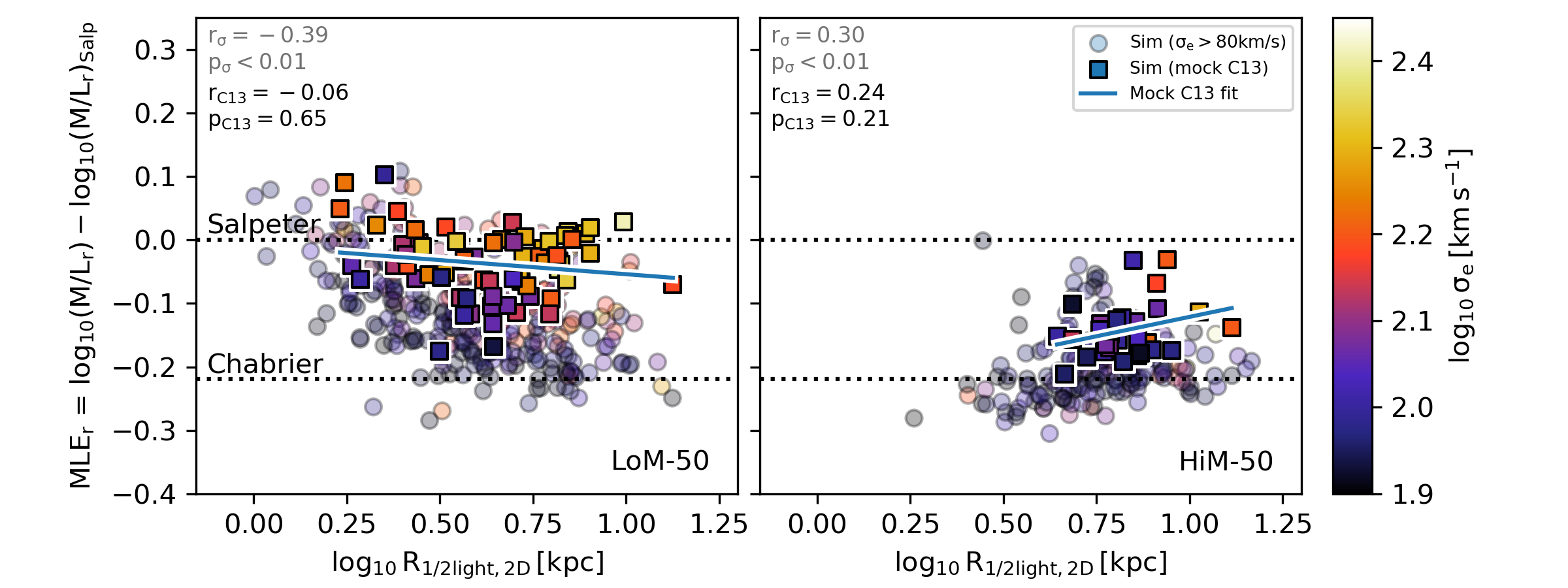}

\caption{As \Fig{IMF_vs_global_props} but now showing $M/L_r$ excess as a function of, from top to bottom, stellar mass reinterpreted assuming a Chabrier IMF, K-band absolute magnitude, and projected $r$-band half-light radius. For completeness we include all galaxies with $\sigma_e > 10^{1.6} (\approx 40)\kms$ in the upper two rows, while the lower row shows galaxies with $\sigma_e > 10^{1.9} (\approx 80)\kms$. We see roughly the same trends of MLE$_r$ with $\MstarChab$ and $M_K$ as with $\sigma_e$ (see Fig. 5 of Paper I), but with greater scatter. In \lom{}, smaller high-$\sigma_e$ galaxies tend to form the more bottom-heavy populations, while larger counterparts form the more top-heavy populations in \him{}. }
\label{fig:MLE_vs_mass_size}
\end{figure*}

\begin{figure*}
\includegraphics[width=\textwidth]{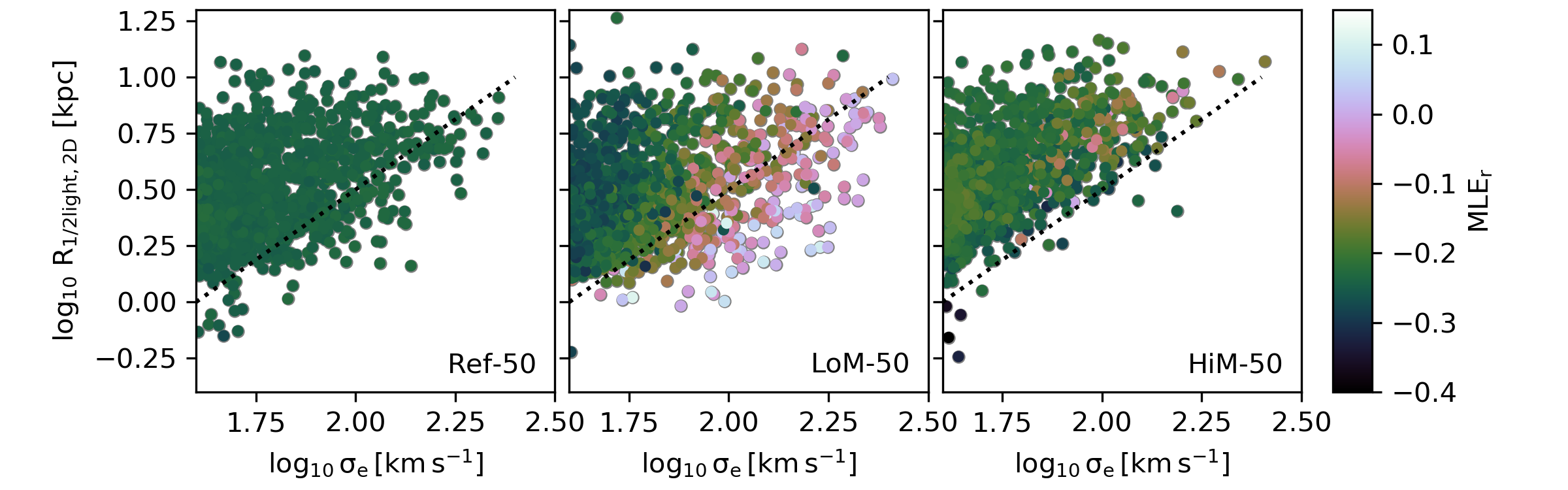}
\caption{Effect of IMF variations on the relation between $r$-band half-light radius and $\sigma_e$. We show all galaxies with $\sigma_e > 10^{1.6}\kms$ at $z=0.1$ in, from left to right, Ref-50, \lom{} and \him{}, respectively. Points are coloured by MLE$_r$. The same (arbitrary) dotted line is repeated in each panel to guide the eye. Galaxies that are smaller at fixed $\sigma_e$ have larger MLE$_r$ in \lom{}. In \him{}, the $r_e-\sigma_e$ relation is tighter, likely due to the stronger feedback in high-pressure (i.e. top-heavy IMF) environments.}
\label{fig:Re_vs_veldisp}
\end{figure*}

In order to build intuition on how the IMF varies from galaxy to galaxy, it is also useful to predict trends between the MLE and other basic galactic properties that have not yet been investigated observationally. In \Fig{MLE_vs_mass_size} we show the MLE$_r$ as a function of, from top to bottom, Chabrier-interpreted stellar mass ($\MstarChab$), $K$-band luminosity, and 2D projected half-light radius for \lom{} (left column) and \him{} (right column) for all galaxies with $\sigma_e > 10^{1.6}$ km s$^{-1}$, coloured by $\sigma_e$.  Note that we now include galaxies of lower $\sigma_e$ than in \Fig{IMF_vs_global_props} to facilitate comparison with the MLE$_r-\sigma_e$ relation shown in Fig. 5 of Paper I, and to show the full transition from Chabrier-like to bottom- or top-heavy IMFs over a wide range of masses and luminosities. Those that would be selected by ATLAS$^{\rm 3D}$ (i.e. are in our mock C13 samples) are shown as opaque squares, while others are shown as translucent circles. $\MstarChab$ is computed by multiplying each galaxy's $K$-band luminosity by the stellar $M/L_K$ that it would have had if its stellar populations had evolved with a Chabrier IMF (given the same ages and metallicities). Note that this is still not exactly the same as would be inferred observationally, as it does not take into account possible biases in the inferred ages or metallicities due to IMF variations. 

The top row shows MLE$_r$ as a function of $\MstarChab$. For \lom{} we see a strong trend of increasing bottom-heaviness with mass for galaxies with $\MstarChab > 10^{10} \Msun$. Galaxies below this limit tend to have Chabrier-like IMFs. The scatter here is stronger than in the MLE$_r-\sigma_e$ relation, leading to a somewhat weaker correlation of MLE$_r$ with $\MstarChab$ for \lom{}. In agreement with \citet{Clauwens2015}, galaxies in our mock C13 sample are only complete down to $\MstarChab \approx 10^{10.5} \Msun$, much higher than the $6\times 10^{9} \Msun$ quoted by \citet{Cappellari2013a}. For \him{}, it is only for the mock C13 galaxies that we see even a weakly positive relation between MLE$_r$ and $\MstarChab$ due to a bias toward high-MLE galaxies at high mass due to the cut in $u^*-r^*$: at fixed mass, galaxies with low MLE tend to be younger, and thus bluer, and are more likely to be excluded from our mock C13 sample. However, in contrast to the significant, positive MLE$_r-\sigma_e$ correlation for this sample, this positive MLE$_r-\MstarChab$ relation for mock C13 \him{} galaxies is not significant, and may be sensitive to the way in which mock C13 galaxies are selected. Note that using true $\Mstar$ on the $x$-axis rather than $\MstarChab$ would shift the highest-MLE galaxies to larger mass by $\approx 0.2-0.3$ dex, and using Salpeter-inferred $\Mstar$ would shift all points systematically to higher mass by 0.22 dex, neither of which would make any difference to these results.

The middle row of \Fig{MLE_vs_mass_size} shows MLE$_r$ as a function of K-band absolute magnitude. For both simulations the trend is very similar to that with $\MstarChab$, but with a shallower slope (in this case the positive relation between the MLE$_r$ and luminosity yields a negative Spearman $r$ for the correlation with magnitude). 

The bottom row shows MLE$_r$ as a function of 2D projected $r$-band half-light radius, $r_e$. In this row we show only galaxies with $\sigma_e > 10^{1.9}\kms$ (as in \Fig{IMF_vs_global_props}) to remove the high number of low-mass galaxies with Chabrier-like IMFs that would reduce the significance of any correlation. Here, the two simulations show markedly different behaviour. While MLE$_r$ decreases with $r_e$ for \lom{}, it increases for \him{}. To help explain this behaviour, we plot in \Fig{Re_vs_veldisp} the $r_e-\sigma_e$ relation for each simulation, coloured by MLE$_r$ (note that this figure contains the same information as the bottom row of \Fig{MLE_vs_mass_size}). For \him{} (right panel), we see a positive correlation between $r_e$ and $\sigma_e$, with no noticeable gradient in MLE$_r$ at fixed $\sigma_e$. Thus, for \him{} the MLE$_r-r_e$ relation matches qualitatively the MLE$_r-\sigma_e$ relation.

In the middle panel of \Fig{Re_vs_veldisp}, we see that, as in \him{}, \lom{} galaxies increase in size with increasing $\sigma_e$, but they also exhibit stronger scatter in the $r_e-\sigma_e$ relation toward smaller galaxies. There is a strong MLE$_r$ gradient at fixed $\sigma_e$ here, where, at fixed $\sigma_e$, smaller galaxies tend to have larger MLE$_r$, resulting in a negative correlation between MLE$_r$ and $r_e$ at fixed $\sigma_e$. This anti-correlation counteracts the (positive) MLE$_r-\sigma_e$ relation, resulting in a net negative correlation between MLE$_r$ and $r_e$ for the $\sigma_e$-complete sample in \lom{} (lower left panel of \Fig{MLE_vs_mass_size}). Interestingly, the trend disappears for the mock C13 sample as many of the smallest (and thus highest MLE$_r$) high-$\sigma_e$ galaxies are excluded due to the luminosity cut, causing these two effects (positive MLE$_r-\sigma_e$ relation coupled with negative MLE$_r-r_e$ at fixed $\sigma_e$) to roughly cancel out.

This behaviour can be explained by the impact of these variable IMF prescriptions on feedback, and thus $r_e$, at fixed $\sigma_e$. In both cases, galaxies that are smaller at fixed $\sigma_e$ tend to have formed their stars at higher pressures, giving them larger MLE$_r$ values. In \lom{}, such galaxies experience weaker feedback due to their bottom-heavy IMFs, further decreasing their sizes relative to galaxies of similar $\sigma_e$ in Ref-50 (compare the middle and left panels of \Fig{Re_vs_veldisp}). The behaviour is different for \him{} due to the fact that galaxies that form at high pressure instead experience stronger feedback due to the top-heavy IMF. This enhanced feedback increases their sizes due to the increased macroscopic efficiency of the ejection of low-angular momentum gas, pushing them upward in the right panel of \Fig{Re_vs_veldisp} (or to the right in the lower right panel of \Fig{MLE_vs_mass_size}). Interestingly, this feedback effect tightens the relationship between $r_e$ and $\sigma_e$ relative to the Ref-50 simulation, resulting in an MLE$_r-r_e$ relation that matches qualitatively the MLE$_r-\sigma_e$ relation.

\subsection{MLE vs $\MBH/\Mstar$}
\label{sec:MLE_vs_MBH_Mstar}

%fig 5: IMF vs MBH/Mstar
\begin{figure*}
\includegraphics[width=\textwidth]{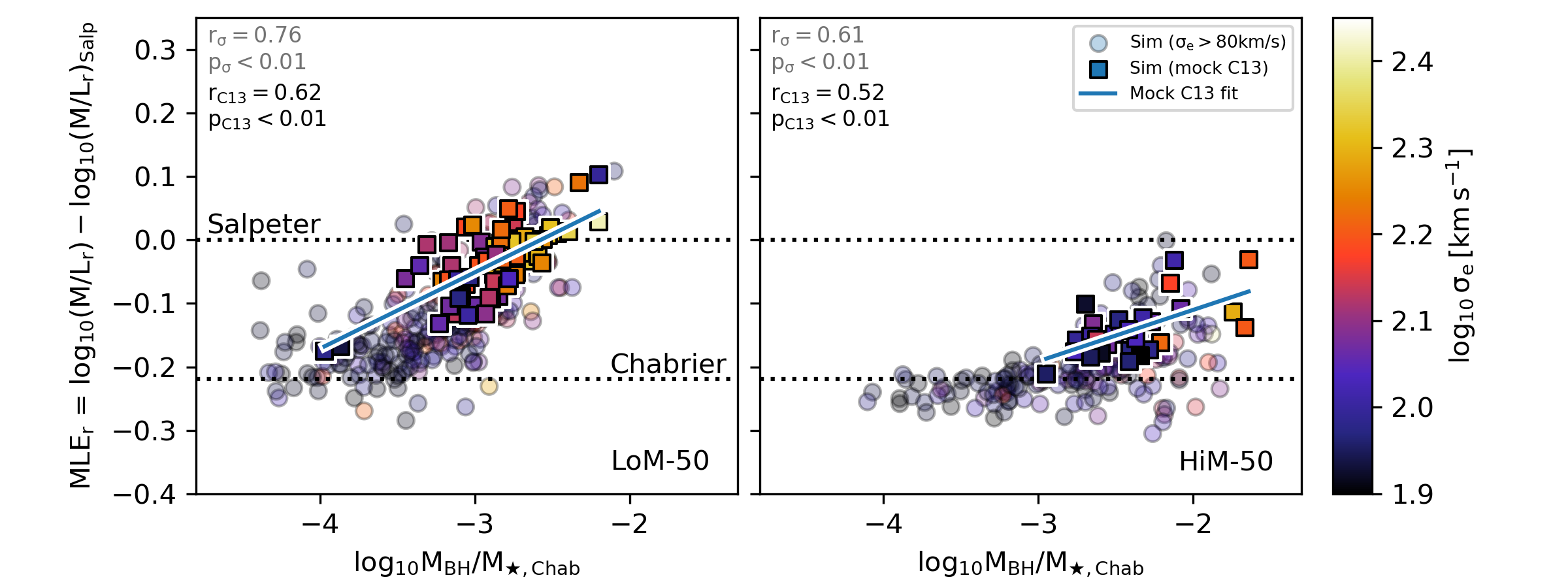}
\caption{As \Fig{IMF_vs_global_props} but now showing the stellar MLE$_r$ as a function of the ratio between $\MBH$ and the stellar mass inferred assuming a Chabrier IMF, $\MstarChab$. In both variable IMF simulations we see a clear trend of increasing MLE$_r$ of the stellar population with higher $\MBH/\MstarChab$. This result shows that an observed correlation between MLE and $\MBH/\Mstar$ does not necessarily imply that $\MBH$ is systematically underestimated for high $\MBH/\MstarChab$ objects. Instead, the trend may be real and a signpost for a variable IMF. The correlations for \lom{} and \him{} come mostly from the fact that galaxies with enhanced $\MBH/\MstarChab$ tend to originate in higher-pressure and older environments, respectively \citep{Barber2016}. }
\label{fig:MBH_Mstar}
\end{figure*}

Some studies \citep[e.g.][]{Pechetti2017} have speculated that the extra dynamical mass inferred in the centres of massive ETGs may be due to an overmassive BH, rather than a variable IMF. For example, NGC 1277 has been found to have an overmassive BH \citep{VanDenBosch2012, Walsh2016} as well as a bottom-heavy IMF \citep{Martin-Navarro2015d}. If central BH masses are computed assuming a stellar population with a Chabrier IMF, while dynamical stellar $M/L$ ratios are inferred assuming central BH masses that lie on the observed $\MBH-\sigma$ relation \citep[e.g.][]{Cappellari2013a}, one would expect a positive correlation between these quantities due to independent studies assigning the excess dynamical mass to either a heavier stellar population or to a massive BH.

It was shown by \citet{Barber2016} that in the reference EAGLE model, which assumes a universal Chabrier IMF, older galaxies tend to have higher BH masses at fixed stellar mass. Since we saw a strong trend between the MLE$_r$ and age in \Sec{IMF_vs_MgFe}, we thus investigate if there is also a trend between the MLE$_r$ and $\MBH/\MstarChab$, shown in \Fig{MBH_Mstar} for our galaxy sample. For both variable IMF simulations, we see a significant positive correlation, where galaxies with high $\MBH/\MstarChab$ also tend to have heavier MLE$_r$. Interestingly, we find a stronger trend of MLE$_r$ with $\MBH/\MstarChab$ for \lom{} than for \him{}. This indicates that the trend is not driven by age, but rather by the fact that the high-pressure environments that lead to the assignment of a heavy IMF also tend to foster the production of overmassive BHs. We have checked that these trends are qualitatively unchanged if using the true $\Mstar$ values on the $x$-axis instead, except with a systematic decrease in $\MBH/\Mstar$ of $\approx 0.1-0.2$ dex. We have thus shown that if the IMF is truly becoming more heavy in high-pressure environments (and hence for higher velocity dispersions), and there is no confusion between the BH mass and stellar mass, we still obtain a positive trend between MLE$_r$ and $\MBH/\Mstar$. Thus, observation of this trend does not necessarily imply that the inferred $M/L$ ratios are increasing due to a systematic underestimate of the BH masses in these high-$\MBH/\Mstar$ galaxies.

The predicted correlation between MLE$_r$ and $\MBH/\MstarChab$ has important implications for the dynamical measurement of BH masses, especially for recently observed galaxies with puzzlingly overmassive BHs. Both of our IMF prescriptions predict a higher {\it stellar} $M/L$ ratio in such galaxies, which must be taken into account when inferring BH masses. Indeed, for our LoM IMF prescription, one would underestimate the stellar mass by a factor of two if one assumes a $M/L$ ratio consistent with a Chabrier IMF when converting the K-band luminosity to $\Mstar$, which would result in an overestimation of $\MBH$. Both the underestimate of $\Mstar$ and the overestimate of the $\MBH$ serve to artificially increase the ratio $\MBH/\MstarChab$. Thus, we suggest that authors who find overmassive black holes in high-mass galaxies consider the possibility that {\it also} the IMF may be either top- or bottom-heavy in these systems.

Indeed, one may turn this argument around and suggest that if one were to seek galaxies with non-Milky Way IMFs, a promising place to look is in galaxies with abnormally high $\MBH/\Mstar$, such as NGC 1277, NGC 1271 \citep{Walsh2015}, and ultra-compact dwarf (UCD) galaxies for which  BH mass measurements are available \citep[e.g. M60-UCD1, ][although tidal stripping may be responsible for the high BH masses in UCDs; \citealt{Barber2016}]{Seth2014}.

\subsection{Which observables correlate most strongly with the MLE?}
\label{sec:most_important_param}

We now determine which of the observable parameters $\sigma_e$, age, [Mg/Fe], metallicity, and $r_e$ best predict MLE$_r$.\footnote{Note that in this analysis we leave out mass and luminosity because these are used to compute MLE in the first place, so correlations between them and the MLE would not be as meaningful.} To do so, we first standardize the logarithm of each parameter such that the mean and dispersion are 0 and 1, respectively, and perform an ordinary linear regression fit to the (standardized) MLE$_r$ using all of these parameters simultaneously. We then determine the change in the adjusted coefficient of determination, $R^2$, when each variable is added to the model last, a quantity we will denote as $\Delta R^2$. Since $R^2$ gives us the fraction of the variance in MLE$_r$ that is explained by the variance in the parameters of the model, $\Delta R^2$ gives us insight into the fraction of the variance in MLE$_r$ that can be accounted for by the variance in each parameter individually, taking into account the information available through the other input parameters. Note that the sum of $\Delta R^2$ values will only equal the total $R^2$ if all of the input variables are completely uncorrelated. For example, two input variables that are strongly correlated will each have $\Delta R^2 \approx 0$, even if individually they can explain much of the scatter in MLE$_r$. Thus, to gain a sense of the contribution of each variable to the total variance in MLE$_r$, we rescale the $\Delta R^2$ values such that they sum to $R^2$.

We first perform this analysis on the mock C13 samples, as they should be more directly comparable to galaxy samples in observational IMF studies. The total $R^2$ values (i.e. the fraction of the variance in MLE$_r$ that can be explained by all of these parameters) are 0.53 and 0.89 for \lom{} and \him{}, respectively.  For \lom{}, we find that $\sigma_e$, $r_e$, and age are the most important variables, with $\Delta R^2 = 0.20$, $0.20$, and $0.14$, respectively, with much smaller contributions from metallicity and [Mg/Fe]. For \him{}, [Mg/Fe] is the most important parameter with $\Delta R^2 = 0.45$, followed by age (0.34) and metallicity (0.12), with negligible contribution from $\sigma_e$ or $r_e$. These results are summarized in \Tab{importance}. 

We conclude that, in a scenario in which the IMF varies at the low-mass end and in such a way as to give the observed trend between MLE$_r$ and $\sigma_e$ (LoM), then when all other variables are kept fixed we obtain the strongest trend of MLE$_r$ with $\sigma_e$, $r_e$, and age, with little to no {\it residual} dependence on metallicity or [Mg/Fe]. The strong trends with $\sigma_e$, $r_e$, and age in our simulations come from the correlation between these parameters and birth ISM pressure, which governs the IMF variations. Indeed, if we include mean light-weighted birth ISM pressure in the fits, the total $R^2$ increases to 0.97, and $\Delta R^2$ drops to $< 0.002$ for all of the other input variables, while that for birth ISM pressure is 0.44.\footnote{Note that $R^2$ is not exactly 1.0 even when correlating MLE with birth ISM pressure directly, due to the fact that the MLE is averaged over many star particles with a potentially wide range of $M/L$ ratios, and that the MLE is not a perfect measure of the IMF even for individual star particles due to some age and metallicity dependence.}

On the other hand, if the IMF varies at the high-mass end (HiM), we find that the MLE$_r$ depends mainly on the age and [Mg/Fe] of the system, with a secondary dependence on metallicity. The strong contribution from [Mg/Fe] comes from the fact that [Mg/Fe] correlates strongly with birth ISM pressure in \him{} galaxies, likely due to the enhanced Mg yields resulting from the IMF becoming top-heavier towards higher pressure environments. Indeed, for the \him{} simulation, the correlation between [Mg/Fe] and birth ISM pressure is stronger than that between $\sigma_e$ and birth ISM pressure, eliminating a need to include $\sigma_e$ in the fit to the MLE when [Mg/Fe] is also included. Remarkably, we find that even adding birth ISM pressure to the list of parameters in the above procedures does not add much new information. In this case, total $R^2$ increases from 0.89 to 0.96, with $\Delta R^2$ = 0.52 and 0.45 for age and birth ISM pressure, respectively, with negligible contributions from the other input variables.  The strong age contribution is due to the dependence of the MLE$_r$ on age when the high-mass slope is shallower than the Salpeter value. Indeed, this age dependence would exist even for a non-variable top-heavy IMF. Note as well that a fit to MLE$_r$ using only birth ISM pressure as an input variable results in $R^2 = 0.97$ and 0.87 for \lom{} and \him{}, respectively (see \Tab{fits}). This result highlights the importance of age on the MLE$_r$ for high-mass, but not low-mass, IMF slope variations.

For completeness, we repeat this analysis for the full sample of galaxies with $\sigma_e > 10^{1.9}\kms$, rather than only those that would have been selected by C13. The results are presented in Tables \ref{tab:fits_sig1p9} and \ref{tab:importance_sig1p9}. For \lom{}, we find qualitatively the same conclusions as for the mock C13 sample, except that now $r_e$ provides much more information than it did for the mock C13 sample (compare Tables \ref{tab:importance} and \ref{tab:importance_sig1p9}), due to the inclusion of compact galaxies with high $\sigma_e$ that are too dim to be included in the mock C13 sample. For \him{}, age, rather than [Mg/Fe], becomes the dominant contributor to the scatter in MLE$_r$ due to the inclusion of young galaxies in the $\sigma_e$-complete sample (see \Fig{MLE_vs_IMFslopes}). These results highlight the importance of sample selection in determining with which property the IMF correlates most strongly.

Finally, we wish to address the question of whether the MLE correlates more strongly with $\sigma_e$ or [Mg/Fe]. Repeating the analysis above for mock C13 galaxies but now only including $\sigma_e$ and [Mg/Fe] in the input parameters, we find that for \lom{}, we obtain $R^2 = 0.19$, with $\Delta R^2 = 0.21$ and $-0.02$ for $\sigma_e$ and [Mg/Fe], respectively\footnote{Note that since we use the adjusted $R^2$, it is possible to have slightly negative $\Delta R^2$ for some variables due to the penalty for adding extra data with no extra information.}. The opposite is true for \him{}, where we obtain $R^2 = 0.70$, and $\Delta R^2 = -0.01$ and $0.71$ for $\sigma_e$ and [Mg/Fe], respectively. Thus, at fixed $\sigma_e$ no correlation exists between MLE$_r$ and [Mg/Fe] for low-mass slope variations, while for high-mass slope variations, no correlation exists between MLE$_r$ and $\sigma_e$ at fixed [Mg/Fe]. These differences are due to the fact that $\sigma_e$ correlates more strongly than [Mg/Fe] with birth ISM pressure in \lom{}, but the opposite is true in \him{} due to the enhanced Mg yields resulting from a top-heavy IMF in high-pressure environments.

It will be interesting to see on which variables the MLE depends most strongly for observed galaxies, as our results suggest that such relations may be used to break the degeneracy between parametrizations of the IMF. \citet{LaBarbera2015} find that the IMF, when parametrized as a top-light ``bimodal'' IMF, shows no correlation with [Mg/Fe] at fixed $\sigma$, which is consistent with our \lom{} simulation. It would be interesting to know if this is still true when parametrizing the IMF with low-mass slope variations instead (as in LoM). Indeed, \citet{Conroy2012b} find that the IMF correlates more strongly with [Mg/Fe] than with $\sigma$ when varying the low-mass slope of the IMF. However, comparisons between IMF studies are difficult due to differences in apertures, methods, and IMF parametrizations employed. We encourage spectroscopic IMF studies to test different parametrizations of the IMF to assess the robustness of correlations between the IMF and galaxy properties, which can then be compared with the predictions presented here. Note that this does not apply to studies that measure the MLE dynamically, as no assumption of IMF parametrization is needed.

\begin{table}
\caption{Determination of the importance of different observables for predicting MLE$_r$ for mock C13 galaxies in our variable IMF simulations. Column 2 gives the fraction of the variance in MLE$_r$ that is accounted for by the variance in each variable indicated in Column 1 (see text).}
\centering
\begin{tabular}{lr} $x$ & {$\Delta R^2$} \\
\hline
\lom{}\\
\hline
$\log_{10}\sigma_e/{\rm (\kms)}$ & $0.20$ \\
$\log_{10} {\rm Age/Gyr}$ & $0.14$ \\
$\log_{10} Z/Z_\odot$ & $< 0.01$ \\
$[$Mg/Fe$]$ & $< 0.01$ \\
$\log_{10} R_e/{\rm kpc}$ & $0.20$ \\
\hline
\him{}\\
\hline
$\log_{10}\sigma_e/{\rm (\kms)}$ & $< 0.01$ \\
$\log_{10} {\rm Age/Gyr}$ & $0.34$ \\
$\log_{10} Z/Z_\odot$ & $0.12$ \\
$[$Mg/Fe$]$ & $0.45$ \\
$\log_{10} R_e/{\rm kpc}$ & $< 0.01$ \\

\end{tabular}
\label{tab:importance}
\end{table}

\begin{table}
\caption{As in \Tab{importance} but for all galaxies with $\sigma_e > 10^{1.9}\kms$. Relative to the mock C13 sample, the importance of $r_e$ and age are enhanced for \lom{} and \him{}, respectively, due to in inclusion of dim, compact galaxies with high MLE$_r$ in the former, and young, low-MLE$_r$ galaxies in the latter. }
\centering
\begin{tabular}{lr} $x$ & {$\Delta R^2$} \\
\hline
\lom{}\\
\hline
$\log_{10}\sigma_e/{\rm (\kms)}$ & $0.22$ \\
$\log_{10} {\rm Age/Gyr}$ & $0.16$ \\
$\log_{10} Z/Z_\odot$ & $< 0.01$ \\
$[$Mg/Fe$]$ & $< 0.01$ \\
$\log_{10} R_e/{\rm kpc}$ & $0.41$ \\
\hline
\him{}\\
\hline
$\log_{10}\sigma_e/{\rm (\kms)}$ & $< 0.01$ \\
$\log_{10} {\rm Age/Gyr}$ & $0.69$ \\
$\log_{10} Z/Z_\odot$ & $0.06$ \\
$[$Mg/Fe$]$ & $0.16$ \\
$\log_{10} R_e/{\rm kpc}$ & $< 0.01$ \\
\end{tabular}
\label{tab:importance_sig1p9}
\end{table}

\section{ MLE of satellite galaxies }
\label{sec:satellites}
\begin{figure*}
   \centering

\includegraphics[width=0.49\textwidth]{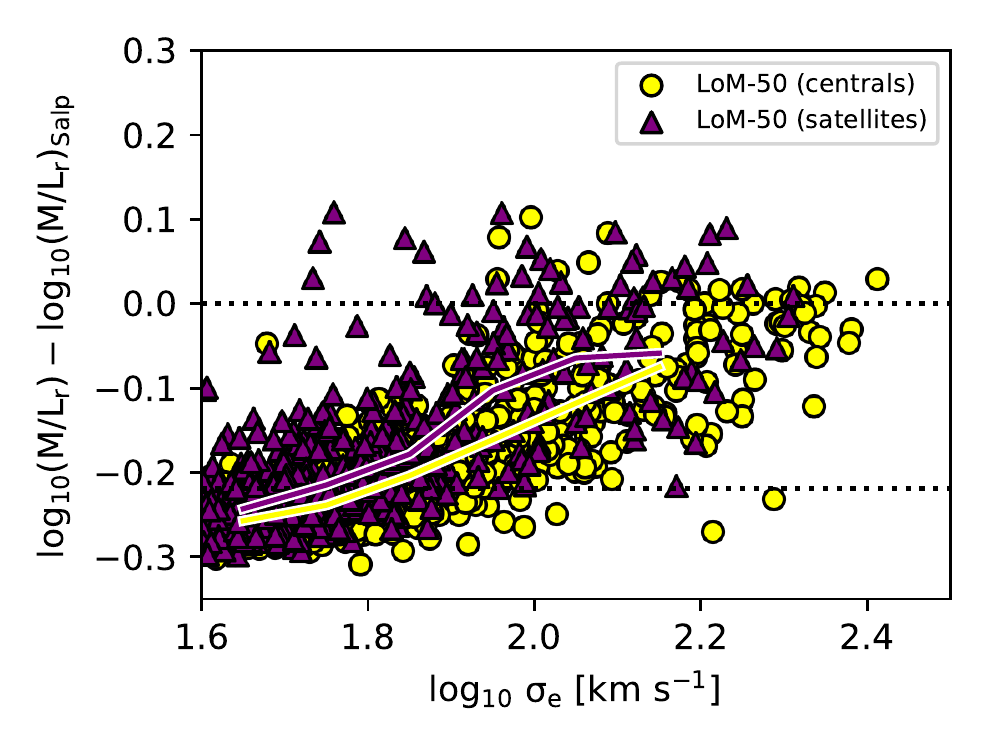}
\includegraphics[width=0.49\textwidth]{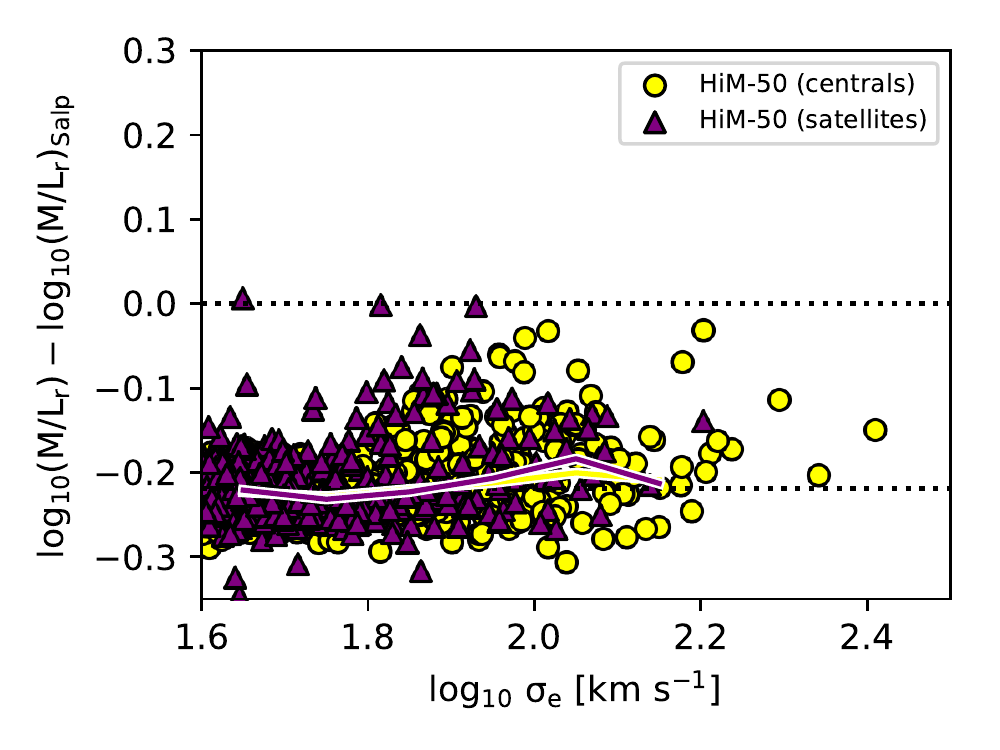}
 \caption{Excess $M/L_r$-ratio relative to a Salpeter IMF as a function of the central stellar velocity dispersion, $\sigma_e$, separated into central (yellow circles) and satellite (purple triangles) galaxies at $z=0.1$. The left and right panels show galaxies from \lom{} and \him{}, respectively. All quantities are measured within the 2D projected stellar $r$-band half-light radius. Solid lines indicate running medians of bin size 0.1 dex in $\sigma_e$. Satellites generally follow the same trend as centrals, but lower-$\sigma_e$ satellites scatter toward higher MLE$_r$ than centrals due to tidal stripping leaving only the IMF-heavy, inner regions bound to the subhalo. }
  \label{fig:IMF_vs_sigma_sats_cents}
 \end{figure*}

We also briefly investigate the effect of environment on the IMF, where in \Fig{IMF_vs_sigma_sats_cents} we show the MLE$_r-\sigma_e$ relation at $z=0.1$ for the two variable IMF simulations split into central and satellite galaxies. Central galaxies are defined for each FoF group as the subhalo to which the most bound gas particle of the group is bound, while all other subhaloes within the group are satellites. For both simulations, on average there is little difference between the two populations, given their significant overlap at fixed $\sigma_e$. However, some satellites tend to scatter toward higher MLE$_r$-values than centrals, especially for $\sigma_e < 100\kms$. This effect is stronger in \lom{} (although it is still visible in \him{}), resulting in a median MLE$_r$ for satellites that is larger by $\approx 0.05\, {\rm dex}$ at $\sigma \approx 100 \kms$. These outliers are likely the stellar cores left over from tidal stripping events. This result can be understood in the context of radial IMF variations, where the central, more tightly bound stars, being born at higher pressures than those in the outer regions, have heavier IMFs than the outer regions that have since been stripped. Indeed, we will show in Paper III that such radial IMF gradients are stronger in \lom{} galaxies than in \him{}.

This result has important consequences for the inference of the stellar mass of satellite galaxies, where, for these variable IMF prescriptions, the underestimate of the inferred stellar masses (assuming a Chabrier IMF) may be as high as a factor of two if they have been significantly stripped. Indeed, recent studies \citep[e.g.][]{Mieske2013, Seth2014, Villaume2017b} find elevated dynamical $M/L$ ratios in UCDs, and have argued for one of two scenarios: either i) these UCDs are the remnant cores of tidally stripped progenitor galaxies and the extra mass comes from a now overmassive central BH, or ii) the IMF in these galaxies is top- or bottom-heavy.  Our results show that both cases can be expected to occur simultaneously, as tidal stripping will increase both MLE$_r$, as the ``IMF-lighter'' outskirts are stripped, as well as $\MBH/\Mstar$, as the galaxy loses stellar mass. 

These results are possibly at odds with the recent work of \citet{Rosani2018}, who found no environmental dependence of the IMF in the $\sigma_e$-stacked spectra of ETGs from SDSS. However, such an analysis would miss outliers due to the $\sigma_e$-stacking. Better statistics at the high-mass end would be required for a more in-depth comparison with their work.

%fig: IMF-sigma for satellites

\section{Summary and Conclusions}
\label{sec:conclusions}

We use cosmological, hydrodynamical simulations with self-consistent variable IMF prescriptions to investigate trends between the mass-to-light excess (MLE) relative to that expected for a Salpeter IMF and global galaxy properties such as age, metallicity, and alpha abundance. These simulations follow the EAGLE reference model \citep{Schaye2015} except that the IMF becomes either bottom-heavy (the \lom{} simulation) or top-heavy (\him{}) for individual star particles formed in high-pressure (or, equivalently, high star formation rate surface density) environments. These simulations are unique in that the IMF variations have been calibrated to match the observed trend of increasing MLE with central stellar velocity dispersion, $\sigma_e$ \citep[][hereafter C13]{Cappellari2013b}. This calibration is possible due to the fact that stars in the centres of ETGs form at higher pressure with increasing $\sigma_e$. The MLE increases with $\sigma_e$ due to an increased fraction of low-mass stars in \lom{}, and an increasing mass fraction of stellar remnants and decreased luminosity in \him{}. We verified in Paper I that the variable IMF simulations reproduce the observables used to calibrate the EAGLE subgrid models for feedback: the galaxy luminosity function, half-light radii and BH masses. The goal of this paper is to determine the similarities and differences in the relationships between the MLE and global galaxy properties that are expected to result from a galaxy formation model with self-consistent, {\it calibrated} IMF variations, in order to help interpret such relationships for observed galaxies and distinguish between different IMF variation scenarios.

Our conclusions are as follows:

\begin{itemize}
\item
The MLE is only a good indicator of IMF slope, independent of age or metallicity, if the high-mass slope is kept fixed at the ``reference'' value (in our case Salpeter). If the high-mass slope is varied, as is the case for \him{}, the MLE becomes as sensitive to age as it is to the IMF (\Fig{MLE_vs_IMFslopes}).
\item
Trends of MLE with overall galactic properties were investigated for properties measured within the half-light radius, $r_e$, for galaxies selected in a way similar to the sample of C13 from ATLAS$^{\rm 3D}$ (\Fig{IMF_vs_global_props}). For \lom{}, MLE correlates positively with age and [Mg/Fe], and negatively with metallicity. However, the significance of the correlations of MLE with metallicity or [Mg/Fe] depends on galaxy selection effects. The anti-correlation with metallicity is the result of the decreasing $Z$ with $\sigma_e$ at high $\sigma_e$ and the tight MLE$-\sigma_e$ relation in that simulation. For \him{}, MLE also correlates positively with age and [Mg/Fe] but has no significant correlation with metallicity due to the fact that the C13 selection criteria exclude faint, old, low-metallicity galaxies with high MLE that would otherwise help drive a negative trend in a $\sigma_e$-complete sample. Fits are shown in \Fig{IMF_vs_global_props} and \Tab{fits}. We will show in a companion paper that the spatially resolved versions of these relations (i.e trends {\it within} galaxies) differ qualitatively from the global trends presented here.
\item
For both \lom{} and \him{}, the MLE correlates positively with mass and luminosity, but with more scatter than the MLE$-\sigma_e$ trend (\Fig{MLE_vs_mass_size}). For \lom{}, MLE anti-correlates with half-light radius, $r_e$, at fixed $\sigma_e$ due to the fact that smaller high-$\sigma_e$ galaxies form stars at higher pressures. This anti-correlation is, however, sensitive to observational selection effects that may prefer larger systems.  MLE correlates positively with $r_e$ for \him{} due to a lack of small high-$\sigma_e$ galaxies, which tightens the positive relation between $r_e$ and $\sigma_e$ in the \him{} simulation (\Fig{Re_vs_veldisp}). These differences can be explained by the weaker (stronger) stellar feedback in high-pressure environments in \lom{} (\him{}) which can decrease (increase) galaxy sizes.
\item A luminosity-complete sample of old, early-type galaxies, as is used in \citet{Cappellari2013b}, exhibits significant biases in the correlations between the MLE and global galaxy properties such as metallicity, stellar mass, luminosity, and size relative to a galaxy sample that is complete in $\sigma_e$ (Figs. \ref{fig:IMF_vs_global_props} and \ref{fig:MLE_vs_mass_size}). Care must thus be taken by observational studies to ensure that selections on properties such as morphology, age, or luminosity do not affect the inferred relations.
\item 
Of the variables $\sigma_e$, age, metallicity, [Mg/Fe], and $r_e$, we determined the importance of each variable in explaining the variance in the MLE. For \lom{}, $\sigma_e$, $r_e$, and age account for the most variance in MLE, while the contributions from metallicity and [Mg/Fe] are much smaller. This result reflects the correlations of birth ISM pressure with $\sigma_e$, $r_e$, and age, and the weak effect that low-mass slope variations have on abundances. For \him{}, [Mg/Fe] and age are the largest contributors to the variance in the MLE, with a smaller but significant contribution from metallicity, and negligible contributions from $\sigma_e$ and $r_e$. This strong age-dependence is due to the sensitivity of the MLE on age for a top-heavy IMF, rather than to any age dependence of the IMF itself. The dependence on abundances likely arises due to the impact of the top-heavy IMF on the stellar yields (see \Tab{importance}).
\item
MLE correlates quite strongly with $\MBH/\MstarChab$ for both variable IMF simulations (\Fig{MBH_Mstar}). This finding suggests that in the variable IMF scenario, galaxies with black holes that are bona fide overmassive relative to the $\MBH-\Mstar$ relation should also have ``heavy'' IMFs. This correlation likely results from the fact that older galaxies tend to form their stars and BHs at higher pressures, leading to both overmassive BHs \citep{Barber2016} and heavy IMFs. We conclude that even though a degeneracy in principle exists between the dynamical calculation of $\MBH$ and stellar $M/L$ ratio in observed galaxies, a correlation between $\MBH/\MstarChab$ and MLE does not imply that overmassive BHs are necessarily the result of incorrect IMF assumptions, nor that excess $M/L$ ratios are solely the result of overmassive BHs. 
\item
Satellite galaxies mainly follow the same MLE$-\sigma_e$ trend as central galaxies, with a few per cent scattering toward high MLE due to tidal stripping removing the outer ``light IMF'' regions, leaving only the ``heavy IMF'' core prior to being completely destroyed by tidal forces. This effect is stronger for \lom{} than for \him{} (\Fig{IMF_vs_sigma_sats_cents}).\\

\end{itemize}

Overall, we have found that an IMF that varies with the local physical conditions, in our case birth ISM pressure, yields a galaxy population for which the MLE correlates with many global galaxy properties. Interestingly, if the high-mass end of the IMF varies, then some correlations can be as strong or stronger than the correlation with birth pressure due to the strong dependence of MLE on age. Our two IMF prescriptions yield qualitatively similar correlations between the MLE and global galaxy properties, particularly with age and [Mg/Fe]. The difference between their predicted correlations with $r_e$ as well as that between the importance of different parameters in predicting the MLE, can be used to differentiate between these two IMF variation scenarios. In Paper III we will investigate further the predicted differences between these simulations by probing the spatially-resolved inner regions of high-mass galaxies to uncover the local behaviour of IMF variations and their impact on, and scaling with, other spatially-resolved properties. Paper III will additionally investigate the effect of these IMF variations on the evolution of galaxies in the simulations.

\section*{Acknowledgements}

 We are grateful to the anonymous referee for helping to improve the quality of the paper. This work used the DiRAC Data Centric system at Durham University, operated by the Institute for Computational Cosmology on behalf of the STFC DiRAC HPC Facility (www.dirac.ac.uk). This equipment was funded by BIS National E-infrastructure capital grant ST/K00042X/1, STFC capital grants ST/H008519/1 and ST/K00087X/1, STFC DiRAC Operations grant ST/K003267/1 and Durham University. DiRAC is part of the National E-Infrastructure. RAC is a Royal Society University Research Fellows. We also gratefully acknowledge PRACE for awarding us access to the resource Curie based in France at Tr$\grave{\rm e}$s Grand Centre de Calcul. This work was sponsored by the Dutch National Computing Facilities Foundation (NCF) for the use of supercomputer facilities, with financial support from the Netherlands Organization for Scientific Research (NWO). This research made use of {\sc astropy}, a community-developed core {\sc python} package for Astronomy \citep{Astropy2013}.
%%%%%%%%%%%%%%%%%%%%%%%%%%%%%%%%%%%%%%%%%%%%%%%%%%



\bibliographystyle{mnras} % This defines the style of the bibliography
\bibliography{IMF} % This is the file with all the references

%%%%%%%%%%%%%%%%%%%%%
%appendix figures
%%%%%%%%%%%%%%%%%%%%%

%%%%%%%%%%%%%%%%%%%%%%%%%%%%%%%%%%%%%%%%%%%%%%%%%%

% Don't change these lines
\bsp	% typesetting comment
\label{lastpage}
\end{document}